\documentclass[longauth]{aa}
\usepackage{graphicx}
\usepackage{txfonts}

\usepackage{natbib}
\bibpunct{(}{)}{;}{a}{}{,}
\usepackage{epsf}
\usepackage{lscape}

\newcommand{\bdv}[1]{\mbox{\boldmath$#1$}}
\def\bpi{{\bdv{\pi}}}
\def\bxi{{\bdv{\xi}}}
\def\bmu{{\bdv{\mu}}}

\begin{document}
\onecolumn{

\title{Mass measurement of a single unseen star\\ and planetary detection efficiency for\\ OGLE 2007-BLG-050}   
\titlerunning{OGLE-2007-BLG-050 : Planetary detection efficiency}
\authorrunning{V.~Batista \emph{et al.}}
\author{V. Batista\inst{1,51}, Subo Dong\inst{2,52}, A. Gould\inst{2,52}, J.P. Beaulieu\inst{1,51},  A. Cassan\inst{3,51}, G.W. Christie\inst{25,52}, C. Han\inst{30,52}, A. Udalski\inst{24,53}\\
and \\
W. Allen\inst{28}, D.\,L.~DePoy\inst{2}, A. Gal-Yam\inst{29}, B.\,S.~Gaudi\inst{2}, B. Johnson\inst{32}, S. Kaspi\inst{43}, C.\,U.~Lee\inst{35}, D. Maoz\inst{43},
J.~McCormick\inst{33}, I. McGreer\inst{31}, B. Monard\inst{28}, T. Natusch\inst{50}, E. Ofek\inst{34}, B.-G. Park\inst{35}, R.\,W. Pogge\inst{2}, D. Polishook\inst{43}, A. Shporer\inst{43} \\
({The $\mu$FUN Collaboration}),\\
M.D.~Albrow\inst{5}, D.~P.~Bennett\inst{4,54}, S.~Brillant\inst{6}, M.~Bode\inst{7}, D.M.~Bramich\inst{8}, 
M.~Burgdorf\inst{49,50},  J.A.R.~Caldwell\inst{9}, H. Calitz\inst{10}, A. Cole\inst{13}, K. H.~Cook\inst{11}, Ch.~Coutures\inst{12}, 
S.~Dieters\inst{1,13}, M.~Dominik\inst{14}\thanks{Royal Society University Research Fellow}, D.~Dominis Prester\inst{15}, J.~Donatowicz\inst{16},  P.~Fouqu\'e\inst{17}, J.~Greenhill\inst{13},
M.~Hoffman\inst{10}, K.~Horne\inst{14}, U.G.~J{\o}rgensen\inst{18}, N.~Kains\inst{14},
S.~Kane\inst{19},D.~Kubas\inst{1,6}, J.B.~Marquette\inst{1}, R.~Martin\inst{20}, P.~Meintjes\inst{10}, J.~Menzies\inst{21}, K.R.~Pollard\inst{5}, K.C.~Sahu\inst{22}, C.~Snodgrass\inst{6}, I.~Steele\inst{7}, Y.~Tsapras\inst{23}, J.~Wambsganss\inst{3}, A.~Williams\inst{20}, M.~Zub\inst{3}\\
({The PLANET/RoboNet Collaboration}),\\
{\L}. Wyrzykowski\inst{24,27}, M. Kubiak\inst{24}, M.\,K. Szyma{\'n}ski\inst{24}, G. Pietrzy{\'n}ski\inst{24,26}, I. Soszy{\'n}ski\inst{24}, O. Szewczyk\inst{26,24}, K. Ulaczyk\inst{24} \\
({The OGLE Collaboration}), \\
F. Abe\inst{36}, I.\,A.~Bond\inst{37}, A. Fukui\inst{36}, K. Furusawa\inst{36}, J.\,B.~Hearnshaw\inst{5}, S. Holderness\inst{38},
Y. Itow\inst{36}, K. Kamiya\inst{36}, P.\,M. Kilmartin\inst{40}, A. Korpela\inst{40}, W. Lin\inst{37},C.\,H. Ling\inst{37}, K. Masuda\inst{36}, Y. Matsubara\inst{36}, N. Miyake\inst{36},
Y. Muraki\inst{41}, M. Nagaya\inst{36}, 
K. Ohnishi\inst{42}, T. Okumura\inst{36}, Y.\,C. Perrott\inst{46}, N. Rattenbury\inst{46}, To. Saito\inst{44}, T. Sako\inst{36}, L. Skuljan\inst{37}, D. Sullivan\inst{39}, 
T. Sumi\inst{36}, W.\,L. Sweatman\inst{37}, P.\,J. Tristram\inst{40}, P.\,C.\,M. Yock\inst{46} \\
({The MOA Collaboration})\\
}


\institute{Institut d'Astrophysique de Paris, INSU-CNRS,
98 bis Boulevard Arago, F-75014 Paris, France ; batista, beaulieu and marquett@iap.fr.
\and Department of Astronomy, Ohio State University,
140 W.\ 18th Ave., Columbus, OH 43210, USA ; dong, gould and gaudi@astronomy.ohio-state.edu.
\and Astronomisches Rechen-Institut, Zentrum f\"ur~Astronomie, Heidelberg University, M\"{o}nchhofstr.~12--14, 69120 Heidelberg, Germany
\and University of Notre Dame, Department of Physics, 225 Nieuwland Science Hall, Notre Dame, IN 46556, USA ; bennett@nd.edu.
\and University of Canterbury, Department of Physics \& Astronomy, Private Bag 4800, Christchurch, New Zealand
\and European Southern Observatory, Casilla 19001, Vitacura 19, Santiago, Chile
\and Astrophysics Research Institute, Liverpool John Moores University, Twelve Quays House, Egerton Wharf, Birkenhead CH41 1LD, UK
\and Isaac Newton Group, Apartado de Correos 321, E-38700 Santa Cruz de La Palma, Spain
\and McDonald Observatory, 16120 St Hwy Spur 78, Fort Davis, TX 79734, USA
\and Dept. of Physics / Boyden Observatory, University of the Free State, Bloemfontein 9300, South Africa
\and Lawrence Livermore National Laboratory, IGPP, P.O. Box 808, Livermore, CA 94551, USA
\and DSM/DAPNIA, CEA Saclay, 91191 Gif-sur-Yvette cedex, France
\and University of Tasmania, School of Maths and Physics, Private bag 37, GPO Hobart, Tasmania 7001, Australia
\and SUPA, University of St Andrews, School of Physics \& Astronomy, North Haugh, St Andrews, KY16~9SS, United Kingdom
\and Physics department, Faculty of Arts and Sciences, University of Rijeka, 51000 Rijeka, Croatia
\and Technical University of Vienna, Dept. of Computing, Wiedner Hauptstrasse 10, Vienna, Austria 
\and Observatoire Midi-Pyr\'en\'ees, UMR 5572, 14, avenue Edouard Belin, 31400 Toulouse, France 
\and Niels Bohr Institute, Astronomical Observatory, Juliane Maries Vej 30, DK-2100 Copenhagen, Denmark
\and NASA Exoplanet Science Institute, Caltech, MS 100-22, 770 South Wilson Avenue Pasadena, CA 91125, USA
\and Perth Observatory, Walnut Road, Bickley, Perth 6076, Australia
\and South African Astronomical Observatory, P.O. Box 9 Observatory 7935, South Africa
\and Space Telescope Science Institute, 3700 San Martin Drive, Baltimore, MD 21218, USA 
\and Astronomy Unit, School of Mathematical Sciences, Queen Mary, University of London, Mile End Road, London E1 4NS, UK
\and Warsaw University Observatory, Al.~Ujazdowskie~4, 00-478~Warszawa,
Poland; udalski, mj, msz, mk, pietrzyn, soszynsk, kulaczyk@astrouw.edu.pl.
\and Auckland Observatory, Auckland, New Zealand; gwchristie@christie.org.nz.
\and Universidad de Concepci{\'o}n, Departamento de Fisica,
Casilla 160--C, Concepci{\'o}n, Chile; szewczyk@astro-udec.cl.
\and Institute of Astronomy, University of Cambridge, Madingley Road,
Cambridge CB3 0HA, UK; wyrzykow@ast.cam.ac.uk.
\and Bronberg Observatory, Centre for Backyard Astrophysics Pretoria, South Africa; lagmonar@nmsa.org.
\and Benoziyo Center for Astrophysics, Weizmann Institute of Science,
76100 Rehovot, Israel; avishay.gal-yam@weizmann.ac.il.
\and Program of Brain Korea, Department of Physics, 
Chungbuk National University, 410 Seongbong-Rho, Hungduk-Gu, 
Chongju 371-763, Korea; cheongho@astroph.chungbuk.ac.kr.
\and Department of Astronomy, Columbia University, Pupin Physics Laboratories, New York, NY 10027; mcgreer@astro.columbia.edu.
\and Institute of Astronomy, Cambridge University, Madingley Rd., Cambridge, CB 0HA, UK; bjohnson@ast.cam.ac.uk.
\and Farm Cove Observatory, Centre for Backyard Astrophysics,
Pakuranga, Auckland New Zealand; farmcoveobs@xtra.co.nz.
\and Division of Physics, Mathematics and Astronomy, California 
Institute of Technology, Pasadena, CA 91125; eran@astro.caltech.edu.
\and Korea Astronomy and Space Science Institute, 61-1 
Hwaam-Dong, Yuseong-Gu, Daejeon 305-348, Korea; bgpark@kasi.re.kr.
\and Solar-Terrestrial Environment Laboratory, Nagoya University, 
Nagoya, 464-8601, Japan.
\and Institute for Information and Mathematical Sciences, Massey University, 
Private Bag 102-904, Auckland 1330, New Zealand.
\and Computer Science Department, University of Auckland, Auckland, New Zealand.
\and School of Chemical and Physical Sciences, Victoria University, 
Wellington, New Zealand.
\and Mt. John Observatory, P.O. Box 56, Lake Tekapo 8770, New Zealand.
\and Department of Physics, Konan University, Nishiokamoto 8-9-1, 
Kobe 658-8501, Japan.
\and Nagano National College of Technology, Nagano 381-8550, Japan.
\and Wise Observatory, Tel Aviv University, 69978 Tel Aviv, Israel; dani, shai, david and shporer@wise.tau.ac.il.
\and Tokyo Metropolitan College of Industrial Technology, Tokyo 116-8523, Japan.
\and Department of Physics and Astrophysics, Faculty of Science, Nagoya University, Nagoya 464-8602, Japan.
\and Department of Physics, University of Auckland, Private Bag 92-019, 
Auckland 1001, New Zealand.
\and Alvine Estate, 456D Vintage Lane, RD3, NZ Blenheim 7321.
\and Deutsches SOFIA Institut, Universitat Stuttgart, Pkaffenwaldring 31, 70569 Stuttgart.
\and SOFIA Science Center, Mail stop N211-3, Moffett Field CA 94035, USA.
\and AUT University, Auckland, New Zealand; tim.natush@aut.ac.nz.
\and Probing Lensing Anomalies NETwork (PLANET).
\and Microlensing Follow Up Network ($\mu$FUN).
\and Optical Gravitational Lens Experiment (OGLE).
\and Microlensing Observations in Astrophysics (MOA).
}

\date{Submitted}
\abstract 
{}
{We analyze OGLE-2007-BLG-050, a high magnification microlensing event ($A\sim 432$) whose peak occurred on 2 May, 2007, with pronounced finite-source and parallax effects. We compute planet detection efficiencies for this event in order to determine its sensitivity to the presence of planets around the lens star.}
{Both finite-source and parallax effects permit a measurement of the angular Einstein radius $\theta_{\rm E}=0.48\pm 0.01$ mas and the parallax $\pi_{\rm E}=0.12\pm 0.03$, leading to an estimate of the lens mass $M=0.50\pm0.14\,M_{\odot}$ and its distance to the observer $D_L=5.5\pm0.4$ \rm{kpc}. This is only the second determination of a reasonably precise ($<30\%$) mass estimate for an isolated unseen object, using any method. This allows us to calculate the planetary detection efficiency in physical units $(r_\perp,m_p)$, where $r_\perp$ is the projected planet-star separation and $m_p$ is the planet mass.}
{When computing planet detection efficiency, we did not find any planetary signature and our detection efficiency results reveal significant sensitivity to Neptune-mass planets, and to a lesser extent Earth-mass planets in some configurations. Indeed, Jupiter and Neptune-mass planets are excluded with a high confidence for a large projected separation range between the planet and the lens star, respectively [0.6 - 10] and [1.4 - 4] AU, and Earth-mass planets are excluded with a 10\% confidence in the lensing zone, i.e. [1.8 - 3.1] AU.}
{}
\keywords{extrasolar planets - gravitational microlensing}

\maketitle


\section{Introduction}
Over the last decade, microlensing events have been intensively followed in order to detect extrasolar planets around lens stars and to measure their abundance in our Galaxy. This is one of the few planet-detection techniques that is sensitive to very low mass planets, and microlensing discoveries have twice the record for the lowest mass planet to orbit a star other than a stellar remnant \citep{beaulieu06,bennett08}. During a microlensing event, i.e. when a background source passes close to the line of sight to a foreground lens star, the observed source flux is magnified by the gravitational field of the lens. The presence of a companion around the lens star introduces two kinds of caustics into the magnification pattern : one or two ``planetary caustics'' associated with the planet and a ``central caustic'' close to the primary lens projected on the source plane. When the source crosses or approaches one of these features, deviations appear from a single point-lens light curve  \citep{maopacz91,gouldloeb92}.

\subsection{Central caustic and detection efficiency}
Significant effort has been expended on the observation and modeling of high magnification events because they probe the central caustic \citep{griestsaf98,rhie00,ratten02}. Any planets in the system are highly likely to affect the central caustic, resulting in potentially high sensitivity to the presence of even low-mass planets.

Indeed, a major advantage of the central caustic is that it is possible to predict in advance when the source passes close to the line of sight of the lens and so when there is the greatest chance of detecting planets. Thus observations can be intensified, further improving the sensitivity to planetary-induced anomalies in the lightcurve.

In these specific cases, for which the impact parameter can be very small, finite-source effects might strongly affect and diminish a possible planetary signal (e.g., \citealt{dong08b}, \citealt{bennettrhie96}). In the absence of any deviation from a finite-source single point-lens model, one can still compute the planet detection efficiency in order to derive upper limits on the probability that the lens harbors a planet \citep{gaudisack00}. It also allows to combine statistically the detection efficiencies computed from observed events to estimate the frequency of planetary companions to the lens \citep{gaudi02}.

The extremely high magnification microlensing event OGLE-2007-BLG-050 was well followed and is a good candidate for analyzing the sensitivity of such an event with pronounced finite-source effects to the presence of a planetary companion. In this study, we compute the planetary detection efficiency for this event, following the \cite{
gaudisack00} method. To perform the calculations of binary light curves, we use the binary-lens finite-source algorithm developed by \cite{dong06} and the formalism of \cite{yoo04} for the single-lens finite-source effects.

\subsection{Mass and distance estimates of the lens star}

OGLE-2007-BLE-050 is also one of the rare events that can potentially be completely solved by measuring both the microlens Einstein angular radius $\theta_{\rm E}$ and the microlens parallax $\pi_{\rm E}$. Indeed, after the first microlenses were detected \citep{alcock93,udal93}, several authors showed that the microlens Einstein angular radius $\theta_{\rm E}$,
$$\theta_{\rm E}=\frac{\theta_*}{\rho_*}
\label{eqn:rho2}
$$
could be measured from deviations relative to the standard point-lens \citep{paczyn86} lightcurve, due to finite-source effects \citep{gould94,nemwick94,wittmao94}. The measured parameter associated with these effects is $\rho_*$, corresponding to the angular size of the source $\theta_*$ in units of $\theta_{\rm E}$. The measurement of $\theta_{\rm E}$ constrains the physical properties of the lens and so leads to the first part of a full solution for an event \citep{gould00},
$$\theta_{\rm E}=\sqrt{\kappa M \pi_{\rm rel}},\qquad \kappa\equiv\frac{4G}{c^2AU}\approx8\,{\rm mas}\,M^{-1}_\odot,
\label{eqn:thetaE}
$$
where $M$ is the lens mass and $\pi_{\rm rel}$ is the lens-source relative parallax.
For most events, the only measured parameter that depends on the mass $M$ is the Einstein timescale, $t_{\rm E}$, which is a degenerate combination of the lens mass $M$, the lens-source relative parallax $\pi_{\rm rel}$ and the proper motion $\mu_{\rm rel}$. It can be expressed as :
$$t_{\rm E}=\frac{\theta_{\rm E}}{\mu_{\rm rel}}
\label{eqn:tE}
$$

\cite{gould92} showed that if one measures both $\theta_{\rm E}$ and the microlens parallax, $\pi_{\rm E}$, which is derived from the distortion of the microlens light curve induced by the accelerated motion of the Earth, one can determine
$$\pi_{\rm E}=\sqrt{\frac{\pi_{\rm rel}}{\kappa M}},
\label{eqn:tE2}
$$
and so determine the lens mass and lens-source relative parallax as well,
$$M=\frac{\theta_{\rm E}}{\kappa\pi_{\rm E}}\,;\qquad \pi_{\rm rel}=\pi_{\rm E}\theta_{\rm E}.
\label{eqn:mass}
$$

After thousands of single-lens microlensing events discovered to date, measurements of both $\theta_{\rm E}$ and $\pi_{\rm E}$ still remain a challenge. The microlens parallax $\pi_{\rm E}$ has been measured for more than twenty single lenses (\citealt{alcock95} [the first parallax measurement], \citealt{poin05} and references therein), while the angular Einstein radius $\theta_{\rm E}$ has been measured for only few cases of single lenses (\citealt{alcock97}, 2001 ; \citealt{smithmaowoz03} ; \citealt{yoo04} ; \citealt{jiang04} ; \citealt{cassan06} ; \citealt{gould09}). 

However, reliable mass estimates for isolated stars have been determined with microlensing only twice.
\cite{alcock01} and \cite{gould09} each measured both $\theta_{\rm E}$ and $\pi_{\rm E}$ respectively for MACHO LMC-5 and OGLE 2007-BLG-224. For MACHO LMC-5, good measurements of $\pi_{\rm rel}$ and $\mu_{\rm rel}$ were obtained with the original photometric data and additional high resolution photometry of the lens (HST observations). Only for OGLE 2007-BLG-224 has there been a reliable mass estimate derived using only ground-based photometric data.

All other good microlens stellar mass measurements to date have been obtained for binary (or planetary) lens events : EROS BLG-2000-5 \citep{an02}, OGLE 2006-BLG-109 \citep{gaudi08}, OGLE 2007-BLG-071 \citep{dong09}, OGLE 2003-BLG-267 \citep{jaros05}, OGLE 2002-BLG-069 \citep{kubas05} and OGLE 2003-BLG-235 \citep{bond04}.

\subsection{Detection efficiency in physical units  }

Here, we present ground based photometric data of the event OGLE 2007-BLG-050 which we use, for the first time, to constrain both the presence of planets and the mass of the lens.

This is also the first event for which parallax and xallarap (source orbital motion) are analyzed simultaneously. However, we find that the apparent xallarap signal is probably due to minor remaining systematic effects in the photometry.

Access to the physical properties of the lens allows us to compute the planetary detection efficiency in physical units $(r_\perp,m_p)$, where $r_\perp$ is the projected separation in AU between the planet and the lens and $m_p$ is the planet mass in Earth mass units.

OGLE-2007-BLG-050 had a high sensitivity to planetary companions of the lens, with a substantial efficiency to Neptune-mass planets and even Earth-mass planets.

\section{Observational data}
The microlensing event OGLE-2007-BLG-050 was identified by the OGLE III Early Warning System (EWS ; \citealt{udal03}) ($\alpha = 17h58m19.39s$, $\delta = -28^o 38'59''$ (J2000.0) and $l = +1.67^o$, $b = -2.25^o$) on 2 Mar 2007, from observations carried out with the 1.3 m Warsaw Telescope at the Las Campanas Observatory (Chile). The peak of the event occured on $HJD'\equiv HJD - 2,450,000 = 4221.904$ (2007 May 2 at 9:36 UT).

The event was monitored over the peak by the Microlensing FollowUp Network ($\mu$FUN ; \citealt{yoo04}) from Chile (1.3m SMARTS telescope at the Cerro Tololo InterAmerican Observatory), South Africa (0.35 m telescope at Bronberg observatory), Arizona (2.4 m telescope at MDM observatory, 1.0 m telescope at the Mt Lemmon Observatory), New Zealand (0.40 m and 0.35 m telescopes at Auckland Observatory and Farm Cove observatory respectively) and on the wings from the Vintage Lane (Marlborough, New Zealand), Wise (Mitzpe Ramon, Israel) and Palomar 60-in (Mt Palomar California, USA) observatories. However, the last three were not included in the final analysis because they do not significantly improve the constraints on planetary companions. Data from all the three sites are consistent with single-lens model.

It was also monitored by Microlensing Observations in Astrophysics (MOA) with the 1.8 MOA-II telescope at Mt John University Observatory (New Zealand), and Probing Lensing Anomalies Network (PLANET ; \citealt{albrow98}) from 5 different telescopes : the Danish 1.54 m at ESO La Silla (Chile), the Canopus 1 m at Hobart (Tasmania), the Elizabeth 1 m at the South African Astronomical Observatory (SAAO) at Sutherland, the Rockefeller 1.5 m of the Boyden Observatory at Bloemfontein (South Africa) and the 60 cm of Perth Observatory (Australia).
The RoboNet collaboration also followed the event with their three 2m robotic telescopes : the Faulkes Telescopes North (FTN) and South (FTS) in Hawaii and Australia (Siding Springs Observatory) respectively, and the Liverpool Telescope (LT) on La Palma (Canary Islands).

In this analysis, we use 601 OGLE data points in \textit{I} band, 104 $\mu$FUN data points in \textit{I} band, 77 $\mu$FUN data points close to \textit{R} band, 121 PLANET data points in \textit{I} band, 55 RoboNet data points in \textit{R} band and 239 MOA-Red data points (wide band covering \textit{R} and \textit{I} bands).

\section{Event modelling}

OGLE-2007-BLG-050 is a very high magnification event ($A\simeq 432$) due to its small impact parameter $u_0$. Because they are quite obvious on the observed light curve, finite-source effects must be incorporated in the modeling. Moreover, the long timescale of the event implies that parallax effects are likely to be detectable.

\subsection{Finite-source effects}

When observing a microlensing event, the resulting flux for each observatory-filter $i$ can be expressed as,
$$F_i(t)=F_{s,i}A[u(t)]+F_{b,i},
\label{eqn:flux}
$$
where $F_{s,i}$ is the flux of the unmagnified source, $F_{b,i}$ is the background flux and $u(t)$ is the source-lens projected separation in the lens plane.

When the source can be approximated as a point, the magnification of a single-lens event is given by \citep{einst36,paczyn86}
$$A(u)=\frac{u^2+2}{u\sqrt{u^2+4}}
\label{eqn:mag}
$$
However, in our case the source cannot be considered as a point ($u\lesssim\rho_*$) and the variation in brightness of the source star across its disk must be considered using the formalism of \citet{yoo04}.
When limb-darkening of the source profile are neglected (uniform source), the magnification can be expressed as \citep{gould94a,wittmao94,yoo04},
$$A_{uni}(u/\rho_*)\simeq A(u)B_0(u/\rho_*),\qquad B_0(z)\equiv\frac{4}{\pi}zE(k,z)
\label{eqn:magyoo}
$$
where $E$ is the elliptic integral of the second kind and $k=min(z^{-1},1)$. Separating the $u$ and $z=u/\rho_*$ parameters allows fast computation of extended-source effects.

To include the limb-darkening, we parameterize the source brightness $S$ by,
$$\frac{S(\theta)}{S_0}=1-\Gamma \biggl[1-\frac{3}{2}(1-\cos{\theta}) \biggr],
\label{eqn:bright}
$$
where $\theta$ is the angle between the normal to the stellar surface and the line of sight. The new magnification is then expressed by adding the $B_1(z)$ function of \cite{yoo04} related to the linear limb-darkening law,
$$A_{ld}(u/{\rho_*})=A(u)[B_0(z)-\Gamma B_1(z)].
\label{eqn:ldmagyoo}
$$
The limb-darkening coefficients $\Gamma$ have been taken equal to $0.49$ for the I filter and $0.60$ for the R filter, which are results from a single-lens fit. From \cite{claret00} and \cite{afon00} models, considering a subgiant similar to our source (log g = 4, T = 5250K), we find $0.44$ and $0.53$, respectively for I and R filters. These values are close to those of our model and lead to essentially the same parameter values as shown in TABLE 1.

In Fig.\ref{fig:espl}, we present the OGLE-2007-BLG-050 light curve modeled with extended-source effects (black curve) and without these effects (red curve). Finite-source effects are clearly noticeable by a characteristic flattening and broadening of the light curve at the peak. 

For each data set, the errors were rescaled to make $\chi^2$ per degree of freedom for the best-fit extended-source point-lens (ESPL) model close to unity. We then eliminated the largest outlier and repeated the process until there were no 3 $\sigma$ outliers. None of the outliers constitute systematic deviations that could be potentially due to planets.

\begin{figure}
\centering
\includegraphics[width=6in]{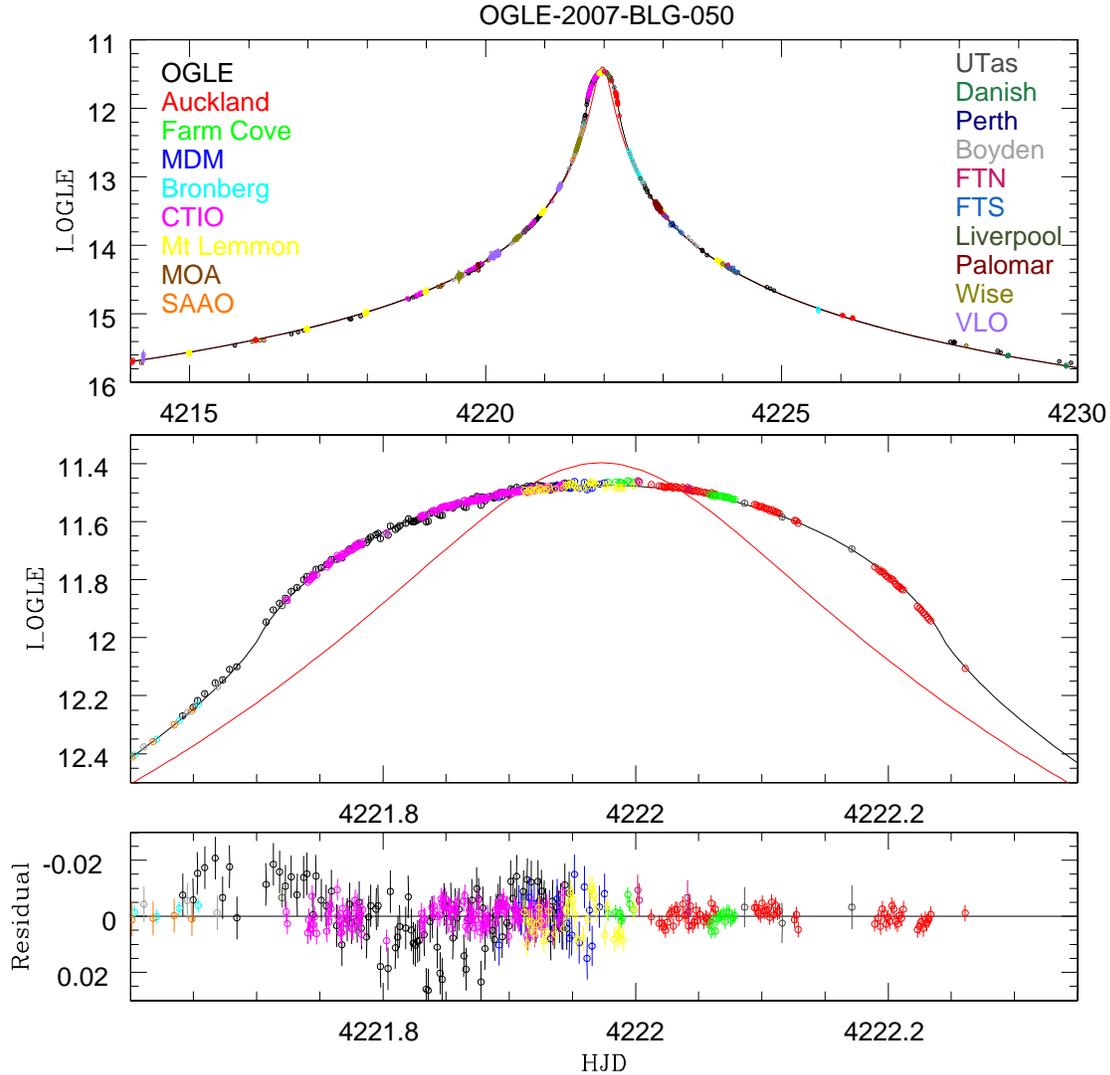}
\caption{\label{fig:espl}
Top: Light curve of OGLE-2007-BLG-050 near its peak on 2007 May 1.
Middle: Zoom onto the peak showing the finite-source effects.
Bottom : Magnitude residuals. They correspond to the real residuals and are not exactly equal to the difference between data and model of the light curve shown above, because the model is given in I band and the R band data points have been linearly converted into the I OGLE system.
We show the model with finite source and parallax effects. As a comparison, a model without finite source effects is shown in red.}\end{figure}

\subsection{Source properties from color-magnitude diagram and measurement of $\theta_{\rm E}$}

To determine the dereddened color and magnitude of the microlensed source, we put the best fit color and magnitude of the source on an  $(I,V-I)$.
calibrated color magnitude diagram (CMD) (cf. Fig.\ref{fig:cmd}). We use calibrated OGLE-III data.
The magnitude and color of the target are $I=18.21\pm 0.03$ and $(V-I)=2.32\pm 0.01$. The mean position of the red clump is represented by an open circle at $(I,V-I)_{RC}=(15.95,2.37)$, with an error of 0.05 for both quantities. The shift in position of our target relative to the red clump is then $\Delta I=2.26\pm 0.05$ and $\Delta (V-I)=-0.05 \pm 0.05$.

For the absolute clump magnitude, we adopt the Hipparcos clump magnitude $M_{I,RC}=-0.23 \pm 0.03$ (\cite{stangarva98}). The mean Hipparcos clump color of $(V-I)_{0,RC}=1.05 \pm 0.05$ is adopted (Jennifer Johnson, 2008, private communication).
Assuming that the source is situated in the bulge and a Galactic center distance of $8 \rm{kpc}$, $\mu_{GC}=14.52\pm0.10$ \citep{eisen05}.

The magnitude of the clump is given by $I_{0,RC}=M_{I,RC}+\mu_{GC}=14.29\pm0.10$. We derive $(I,V-I)_{0,RC}=(14.29,1.05)\pm(0.10,0.05)$. Hence, the dereddened source color and magnitude are given by : $(I,V-I)_0=\Delta(I,V-I)+(I,V-I)_{0,RC}=(16.55,1.00)\pm(0.12,0.08)$.

From $(V-I)_0$, we derive $(V-K)_0$ using the \cite{bessbret88} diagram for giants, supergiants and dwarfs : $(V-K)_0=2.31 \pm 0.13$. The measured values of $I_0$ and $(V-I)_0$ then lead to $K_0=15.24 \pm 0.09$.

For completeness, we also derive an extinction estimate $[A_I,E(V-I)] =(1.66,1.32)$, which leads to an estimate $R_{VI}=A_V/E(V-I)=2.02$.

The color determines the relation between dereddened source flux and angular source radius. We use the following expression given by \cite{kerv04} for giants between A0 and K0 :
$$\log{2\theta_*}=0.5170-0.2K_0+0.0755(V-K)_0,
\label{eqn:source}
$$
giving $\,\theta_*=2.20\pm0.06\,\mu as$.

With the angular size of the source given by the extended source point lens (ESPL) fit, $\rho_*=0.00458\pm0.00003$, we derive the angular Einstein radius $\theta_{\rm E}$ : $\theta_{\rm E}=\theta_*/\rho_*=0.48\pm0.01\,\rm{mas}$, where the error is determined by : $(\sigma_{\theta_{\rm E}}/\theta_{\rm E})^2=(\sigma_{\theta_*}/\theta_*)^2+(\sigma_{\rho_*}/\rho_*)^2$. This first fit takes into account finite source effects only. The values of $\rho_*$ and $\theta_{\rm E}$ will not change significantly when adding new effects (see parallax effects later) but the induced modifications will be included in the final results.

Then, combined with the fitted timescale of the event $t_{\rm E}=66.9\pm0.6\,$days, gives the geocentric relative lens-source proper motion : $\mu=\theta_{\rm E}/t_{\rm E}=2.63\pm0.08\,$mas/yr, with the same method for calculating the error.

\begin{figure}
\begin{center}
\includegraphics[width=4.65in]{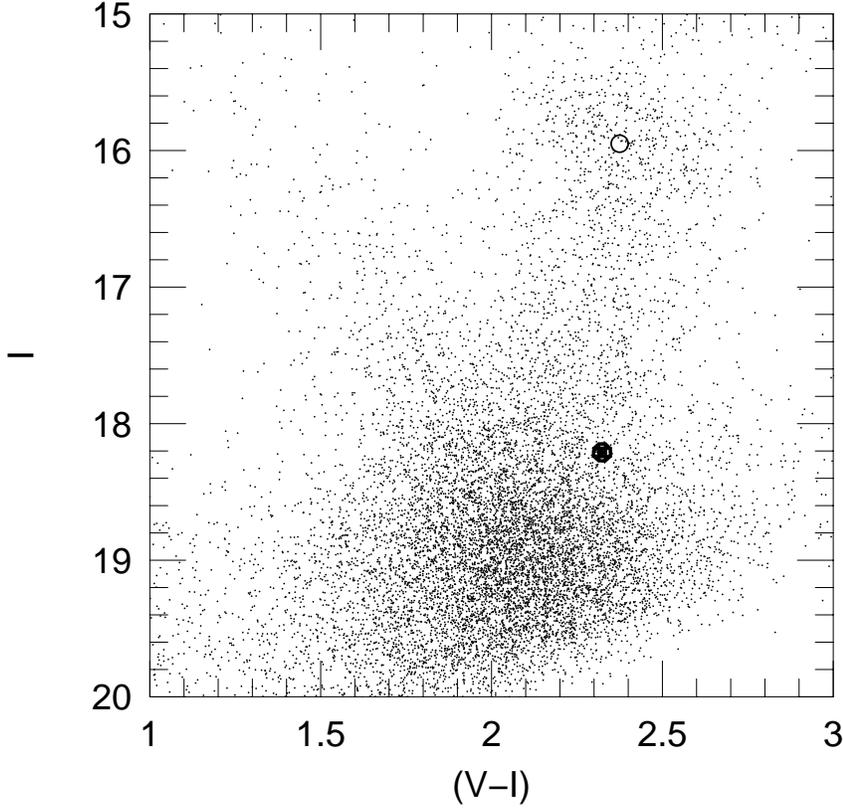}
\caption{Calibrated color-magnitude diagram of the field around OGLE-2007-BLG-050. The clump centroid is shown by an empty open circle, while the OGLE-III $I$ and $V-I$ measurements of the source are shown by an open circle surrounding $1\,\sigma$ error bars.}
\label{fig:cmd}
\end{center}
\end{figure}

\subsection{Parallax effects}

\subsubsection{Orbital parallax effects}

The source-lens projected separation in the lens plane, $u(t)$ of Eq.\ref{eqn:flux}, can be expressed as a combination of two components, $\tau(t)$ and $\beta(t)$, its projections along the direction of lens-source motion and perpendicular to it, respectively :
$$u(t)=\sqrt{\tau^2(t)+\beta^2(t)}.
\label{eqn:ut}
$$
If the motion of the source, lens and observer can all considered rectilinear, the two components of $u(t)$ are given by,
$$\tau(t)=\frac{t-t_0}{t_{\rm E}}\qquad ; \qquad \beta(t)=u_0.
\label{eqn:u1u2}
$$
In the case of a simple point-source point-lens model, only five parameters are fitted : the source flux $F_s$, the blending flux $F_b$ (both duplicated if more than one observatory), the time of the closest approach $t_0$, the impact parameter $u_0$ and the timescale of the event $t_{\rm E}$.

\begin{figure}
\begin{center}
\includegraphics[width=4in]{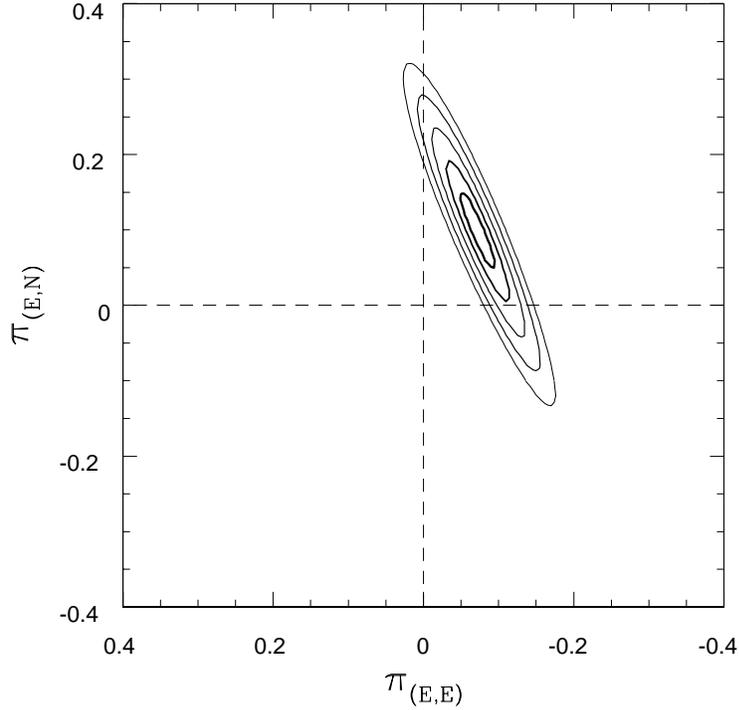}
\caption{\label{fig:parallax}
Likelihood contours as a function of the parallax vector $\bpi_{\rm E}$ ($1,\,2,\,3,\,4\,\sigma$). The best fit is $\bpi_{\rm E}=(0.099,-0.072)$. There is a hard $3\,\sigma$ lower limit $\pi_{\rm E}>0.086$ which implies $M<0.67M_{\odot}$ and a $3\,\sigma$ upper limit $\pi_{\rm E}<0.23$ which implies $M>0.25M_{\odot}$.
}
\end{center}
\end{figure}

However, for long events, like OGLE-2007-BLG-050 (where $t_{\rm E}\geq \rm{yr} /2\pi$), the motion of the Earth cannot be approximated as rectilinear and generates asymmetries in the light curve. Parallax effects then have to be taken into account. To introduce these effects, we use the geocentric formalism (\citealt{an02} and \citealt{gould04}) which ensures that the three standard microlensing parameters ($t_0$, $t_{\rm E}$, $u_0$) are nearly the same as for the no-parallax fit. Now two more parameters are fitted. These are the two components of the parallax vector,  $\bpi_{\rm E}$, whose magnitude gives the projected Einstein radius, $\tilde{r}_{\rm E}=AU/\pi_{\rm E}$ and whose direction is that of lens-source relative motion.

The parallax effects imply additional terms in the Eq.\ref{eqn:u1u2}
$$\tau(t)=\frac{t-t_0}{t_{\rm E}}+\delta \tau(t) \qquad ; \qquad \beta(t)=u_0+\delta \beta(t)
\label{eqn:u1u2bis}
$$
where
$$(\delta \tau(t),\delta \beta(t))=\bpi_{\rm E}\bdv{\Delta s}=(\bpi_{\rm E}.\bdv{\Delta s}, \bpi_{\rm E} \times \bdv{\Delta s)}
\label{paral}
$$
and $\mathbf{\Delta s}$ is the apparent position of the Sun relative to what it would have been assuming a rectilinear motion of the Earth.

The Extended-Source Point-Lens (ESPL) fit yields a determination of the components $(\pi_{E,N},\pi_{E,E})$ of the parallax vector $\bpi_{\rm E}$ projected on the sky in North and East celestial coordinates. This is done by mapping a grid over the $\bpi_{\rm E}$ plane and searching for the minimum of $\chi^2$ (cf. Fig.\ref{fig:parallax}). In addition to the best ESPL fit presented in Section \ref{subsection:fit}, this grid search was done to probe the likelihood contours as a function of $\pi_{\rm E}$, holding each trial parameter pair $\bpi_{\rm E}=(\pi_{E,N},\pi_{E,E})$ fixed while allowing all remaining parameters to vary. The best fit is $\bpi_{\rm E}=(0.099,-0.072)$. There is a hard $3\,\sigma$ lower limit $\pi_{\rm E}>0.086$ and a $3\,\sigma$ upper limit $\pi_{\rm E}<0.23$. The error of $\pi_{\rm E}$ is calculated from the $1\,\sigma$ contour : $\pi_{\rm E}=0.12\pm 0.03$. The likelihood contours in the $\pi_{\rm E}$ plane are slightly elongated along the North-South axis. This tendency, which is weak here due to the long timescale, is explained in \cite{gould94} by the fact that for short events the Earth's acceleration vector is nearly constant during the event.

\begin{figure}
\begin{center}
\includegraphics[width=4in]{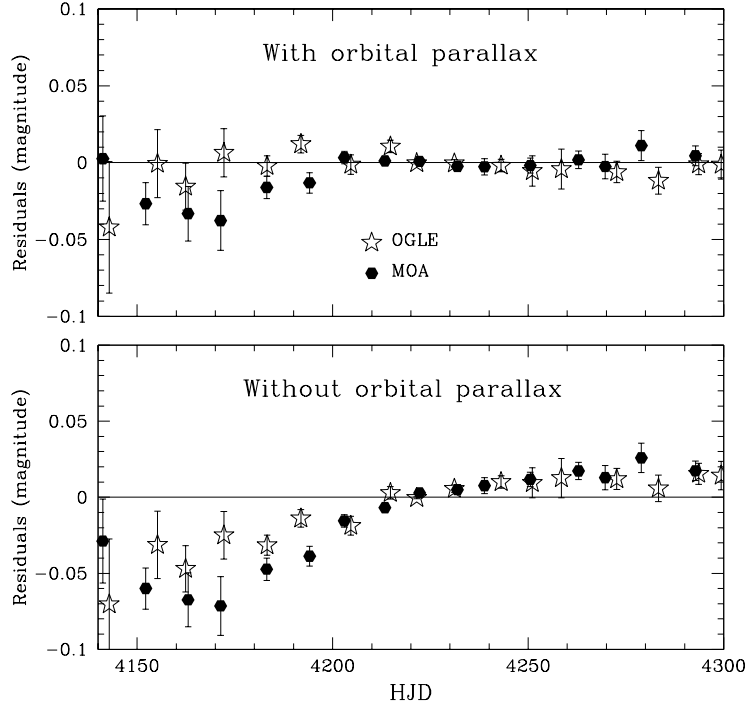}
\caption{\label{fig:residual}
OGLE (stars) and MOA (hexagons) residuals (magnitude) for models with (upper panel) and without (lower panel) parallax effects.
The residuals have been binned for clarity.
}\end{center}
\end{figure}

The Fig.\ref{fig:residual} shows the modeling improvement when we include the orbital parallax effects in the fit. These plots only show the OGLE and MOA residuals because these data mostly constrain the parallax since they cover a long time range.

As discussed by \cite{smithmaopac03}, there is a $u_0 \leftrightarrow -u_0$ degeneracy. For a low magnification event with $\left | u_0 \right | \sim 1$, the $u_0 > 0$ and $u_0 < 0$ solutions will behave differently, but for a high magnification event with $\left | u_0 \right | \ll 1$ like OGLE-2007-BLG-050, the $u_0 \leftrightarrow -u_0$ transformation can be considered as a symmetry and there is no possibility to distinguish one solution from orbital motion alone. In principle, these can be distinguished from so-called ``terrestrial parallax'' effects caused by the different positions of the telescopes on the surface of the Earth.

\subsubsection{Terrestrial parallax effects}
We investigate terrestrial parallax in order to check if it is consistent with the vector parallax determined from orbital parallax effects and to distinguish the $u_0>0$ and $u_0<0$ solutions. The resulting $\chi^2$ of the orbital+terrestrial parallax model does not show any improvement and is actually worse than orbital parallax alone ($\Delta\chi^2=4$, $\chi^2_{orbital\,parallax}=1760.5$). The most likely explanation for this discrepancy is that the much stronger ($\Delta\chi^2=235$, $\chi^2_{without\,parallax}=1995.4$) signal from orbital effects reflects the true parallax and the small terrestrial parallax ``signal'' is actually just due to low-level systematic errors.

\subsubsection{Xallarap effects}
\label{subsubsection:xallarap} 
We also consider the possibility that the orbital parallax signal is actually due to xallarap (orbital motion of the source) rather than to real parallax. Of course an orbital motion of the source, in case of a binary orbit that fortuitously mimics that of the Earth, can reproduce the same light curve as the orbital parallax effects but here we are looking for orbital motion that is inconsistent with the Earth-motion explanation.

We therefore search for xallarap solutions (orbital motion of the source) by introducing 5 new parameters in the model related to the orbital motion of the source : $P$ the period of the source's orbit, $\xi_{E,N}$ and $\xi_{E,E}$ the xallarap vector which is analogous to the $\bpi_{\rm E}$ vector, and $\alpha_2$ and $\delta_2$, the phase and inclination of the binary orbit which function as analogs of the celestial coordinates of the source in case of parallax. The rather long timescale does not justify removing parallax effects to search for xallarap only and moreover, searching for a model including only xallarap effects does not provide significant improvements. For these reasons, we search for a solution that takes into account both orbital + terrestrial parallax and xallarap effects with a Markov Chain Monte Carlo algorithm (MCMC). We explore a large range of periods, from 0 to 700 days, and find a $\chi^2$ improvement ($\chi^2=1717.7$, $\Delta\chi^2=-43$) for periods above 250 days in comparison with the orbital parallax effects only. The $\chi^2$ is essentially flat in the period range [250 - 500] days with a very shallow minimum around $P = 290$ days.

The $P = 290$ days solution gives : $\bxi_{\rm E}=(0.958,-0.273)$, and thus a source orbital radius : $a_s=D_S\,\theta_{\rm E}\,\xi_{\rm E}=3.74\,AU$. 

Kepler's third law (expressed in solar-system units),
$$a^3=P^2M \,; \, M\equiv M_s + M_c
\label{kepler}
$$
and Newton's third law,
$$M_s\,a_s=M_c\,a_c\,\Rightarrow a\equiv a_c + a_s = \bigg(1+\frac{M_s}{M_c}\bigg)a_s
\label{newton}
$$
imply
$$\frac{M^3}{M_c^3}a_s^3=P^2M\,\Rightarrow \frac{a_s^3}{P^2}=\frac{M_c}{[1+(M_s/M_c)]^2}.
\label{companion}
$$
From the position of the source relative to the red clump on the CMD diagram (Fig.\ref{fig:cmd}), we conclude that the source is a sub-giant situated in the bulge and, because the bulge is an old population, we infer that the source mass $M_s$ is close to a solar mass with an upper limit of $1.2 M_{\rm{\odot}}$. 
This mass limit and the long orbital period require a companion with $M_c >70M_{\rm{\odot}}$, thus a black hole, which has an extremely low a priori probability. And if the companion is neither a black hole nor a neutron star, its mass has to be less or equal than the source mass since the source is an evolved star and a slightly heavier companion would therefore be much brighter. To explore these other possible star companions, we add a new constraint on the magnitude of the xallarap vector in the MCMC program, assuming that $M_s < 1.2 M_{\rm{\odot}}$ and $M_c/M_s \le 1$, which can be expressed as : 
$$\xi_E < \frac{0.3^{1/3}}{3.7}P^{2/3}=0.18(P/yr)^{2/3}.
\label{contrainte}
$$

The minimum of $\chi^2$ ($\chi^2 = 1730$) is obtained for a source orbital period equal to 170 days as shown in the Fig.\ref{fig:chi2vsperiod}. When we put the corresponding parameters ($P$, $\alpha_2$, $\delta_2$) in a differential-method program to reach a more accurate solution, we find $\chi^2 = 1728.1$.
\begin{figure}
\begin{center}
\includegraphics[width=4in]{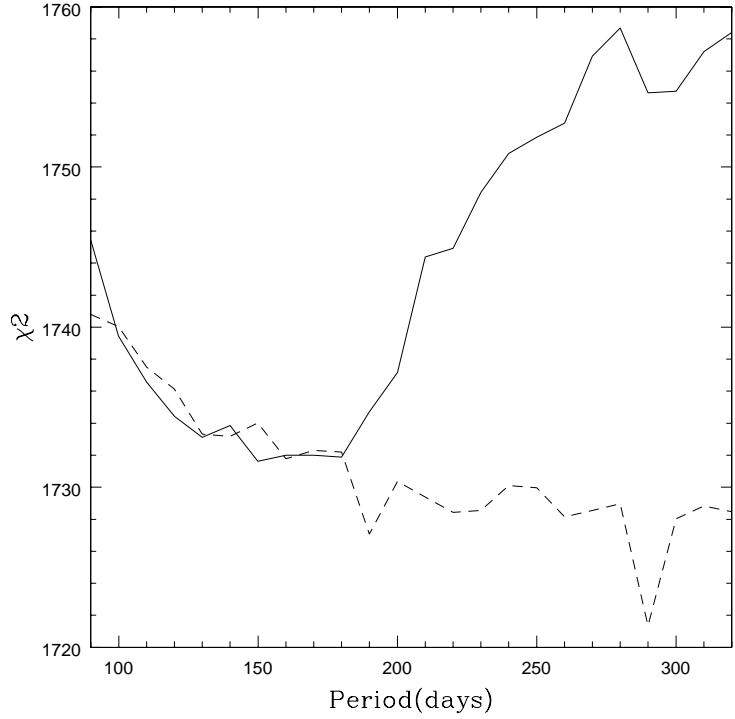}
\caption{
$\chi^2$ as a function of the period of the source'orbit from a MCMC algorithm with parallax and xallarap effects. The dashed line is the case without any constraint on the companion mass and leads to a black hole solution. The solid line is for a constraint ($\xi_E < \frac{0.3^{1/3}}{3.7}P^{2/3}=0.18(P/yr)^{2/3}$) and leads to a solar mass companion with a minimum of $\chi^2$ at $P=170$ days.}
\label{fig:chi2vsperiod}
\end{center}
\end{figure}

The xallarap vector of this solution ($\xi_{E,N}$, $\xi_{E,E}$)=($-0.0142$, $0.0940$) implies a source orbital radius $a_s=D_S\,\theta_{\rm E}\,\xi_{\rm E}=0.40 AU$ and a companion mass close to $1M_{\rm{\odot}}$.
The MCMC algorithm permits us to explore an 11-dimensional space ($t_0$, $t_E$, $u_0$, $\rho_*$, $\pi_{E,N}$, $\pi_{E,E}$, $\xi_{E,N}$, $\xi_{E,E}$, $P$, $\alpha_2$, $\delta_2$). We plot the $1\sigma$ and $3\sigma$ limits of the $\left |\bpi_E \right |=\pi_E$ as given in the Fig.\ref{fig:chi2vspie}.
The resulting parallax is then $\pi_E = 0.94 \pm 0.10$.
\begin{figure}
\begin{center}
\includegraphics[width=4in]{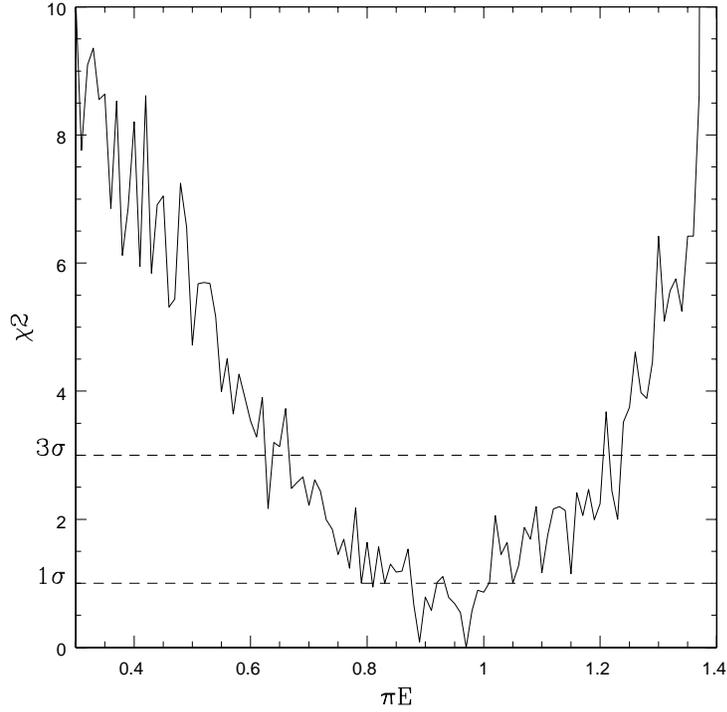}
\caption{
$\chi^2$ as a function of the magnitude of the parallax vector $\bpi_E$ from MCMC runs including the constraint $\xi_E < (0.3^{1/3}/3.7)P^{2/3}=0.18(P/yr)^{2/3}$ on the companion mass. $\pi_E = 0.94 \pm 0.10$.
}
\label{fig:chi2vspie}
\end{center}
\end{figure}

\subsection{Characteristics of the extended-source models with parallax and xallarap effects}
\label{subsection:fit}
Considering the finite-source effects and parallax + xallarap effects, and the 16 observatories involved in the event monitoring, we have to fit 43 parameters (the 3 standard parameters, 1 for the angular size of the source, 2 for parallax, 5 for xallarap and 2x16 for the fluxes $F_s$ and $F_b$ of the different telescopes). The best ESPL fit model including parallax and xallarap effects ($\chi^2 = 1717.7$) corresponds to a binary system in which the source companion is a black hole (see \S.\ref{subsubsection:xallarap}). One more reasonable solution could be the solar mass companion obtained when using a constraint on the xallarap (see \S.\ref{subsubsection:xallarap}). This solution has $\chi^2=1728.1$ for 1745 data points and 43 fit parameters, to give $\chi^2/d.o.f=1.01$, while the best ESPL fit with parallax effects only has $\chi^2=1760.5$ and the one without any parallax nor xallarap effects gives $\chi^2=1995.4$, a difference of $\Delta\chi^2=267.3$.
The corresponding best-fit parameters and their errors as determined from the light curve of three different models are shown in Table 1 and Table 2 (see also Fig.\ref{fig:espl}).

\small{
\begin{table}
\begin{center}
\begin{tabular}{l r r r}
\multicolumn{4}{c}{TABLE 1} \\
\multicolumn{4}{c}{OGLE-2007-BLG-050   \textsc{Fit  Parameters}} \\
\multicolumn{4}{c}{Extended-Source Point-Lens with Parallax} \\
\hline
\hline
\\
\multicolumn{1}{c}{Parameters} & \multicolumn{1}{c}{Without} & \multicolumn{1}{c}{Xallarap} &  \multicolumn{1}{c}{Xallarap}\\
\multicolumn{1}{c}{   } & \multicolumn{1}{c}{xallarap} & \multicolumn{1}{c}{Black hole} & \multicolumn{1}{c}{Solar mass}\\
\multicolumn{1}{c}{   } & \multicolumn{1}{c}{   } & \multicolumn{1}{c}{P = 290 days} & \multicolumn{1}{c}{P = 170 days}\\
\hline
\normalsize{$\chi^2$} &  1760.5 &  1717.7 &  1728.1\\
\hline\\
\normalsize{$t_0\,\,(days)$}  & 4221.9726 &  4221.9725 & 4221.9725\\
\normalsize{$\sigma_{t_0}\,\,(days)$} & 0.0001 &  0.0001 & 0.0001\\
\normalsize{$u_0$} &  0.00204 & 0.00215 &  0.00214\\
\normalsize{$\sigma_{u_0}$} &  0.00002 &  0.00003 &  0.00003 \\
\normalsize{$t_{\rm E}\,\,(days)$}  & 68.09 &  64.96 &  65.11 \\
\normalsize{$\sigma_{t_{\rm E}}\,\,(days)$} & 0.66 &  0.75 &  0.75 \\
\normalsize{$\rho_*$} &  0.00450 &  0.00473 &  0.00471 \\
\normalsize{$\sigma_{\rho_*}$} & 0.00004 & 0.00006 &  0.00006 \\
\normalsize{$\pi_{E}$} &  0.12 &  2.33 &  0.94 \\
\normalsize{$\sigma_{\pi_{E}}$} & 0.03 & 0.07 &  0.10\\
\normalsize{$\xi_{E}$} &  / & 1.00 & 0.17 \\
\normalsize{$\sigma_{\xi_{E}}$} &  / & 0.06 &  0.07\\
\hline
\end{tabular}
\label{tab1}
\end{center}
\caption{\label{table:tab1}
Fit parameters for three different models : 1/ with orbital parallax effects only, 2/ with orbital parallax + xallarap (black-hole source companion), 3/ with orbital parallax + xallarap (solar-mass source companion).
}
\end{table}
}

\begin{table}
\scriptsize{
\begin{center}
\begin{tabular}{l c c c c}
\multicolumn{5}{c}{\small{TABLE 2}} \\
\multicolumn{5}{c}{\small{OGLE-2007-BLG-050 \textsc{Flux Parameters}}} \\
\multicolumn{5}{c}{\small{Extended-Source Point-Lens}}\\
\multicolumn{5}{c}{\small{Orbital Parallax + Xallarap (solar mass companion)}}
\\
\hline
\hline
\\
\textsc{\small{Observatory}} & \small{$F_s$} & \small{$\sigma_{F_s}$} & \small{$F_b$} & \small{$\sigma_{F_b}$} \\
\\
\hline
\\
OGLE \textit{I} & 0.96 & 0.01 & 0.28 & 0.01 \\
MOA \textit{I} & 0.96 & 0.01 & 0.29 & 0.01 \\
$\mu$FUN \textit{R} New-Zealand (Auckland) & 1.08 & 0.01 & 3.06 & 0.05 \\
$\mu$FUN \textit{R} New-Zealand (Farm Cove) & 0.91 & 0.009 & -27.0 & 0.2 \\
$\mu$FUN \textit{I} Arizona (MDM) & 0.99 & 0.17 & 242.3 & 75.0 \\
$\mu$FUN \textit{R} South Africa (Bronberg) & 0.232 & 0.002 & 0.28 & 0.04 \\
$\mu$FUN \textit{I} Chile (CTIO SMARTS) & 6.24 & 0.06 & -3.9 & 0.2 \\
$\mu$FUN \textit{I} Arizona (Mt Lemmon) & 3.94 & 0.04 & 21.8 & 0.1 \\
PLANET \textit{I} South Africa (SAAO) & 5.60 & 0.05 & -1.9 & 1.6 \\
PLANET \textit{I} Australia, Tasmania (UTas) & 2.74 & 0.03 & 163.8 & 1.2 \\
PLANET \textit{I} Chile (Danish) & 14.45 & 0.15 & 13.0 & 0.4 \\
PLANET \textit{I} Australia (Perth) & 0.793 & 0.008 & 6.5 & 0.1 \\
PLANET \textit{I} South Africa (Boyden) & 9.59 & 0.09 & 447.3 & 0.6 \\
Robonet \textit{R} Faulkes North (Hawaii) & 0.115 & 0.001 & -0.26 & 0.02 \\
Robonet \textit{R} Faulkes South (Australia) & 0.129 & 0.001 & -0.229 & 0.009 \\
Robonet \textit{R} Liverpool (Canaries Island) & 2.75 & 0.03 & 17.7 & 0.3\\
\hline
\label{tab2}
\end{tabular}
\end{center}
\small
\caption{\label{table:tab2}
Source flux and blending for telescopes that observed OGLE 07-BLG-050. The given values corresponds to the model with parallax and xallarap, in a case of a solar mass companion for the source. They do not change significantly for the other models.
}}\end{table}

\normalsize

\subsection{Lens mass and distance estimates}
\label{sec:lensmass}

\cite{gould92} showed that if both $\theta_{\rm E}$ and $\pi_{\rm E}$ could be measured, then the mass $M$ and the lens-source relative parallax $\pi_{\rm{\rm rel}}$ could be determined as given in Eq.\ref{eqn:mass} and then the lens distance could be deduced from :
$$\pi_{\rm{\rm rel}}=1 {\rm AU}\biggl(\frac{1}{D_L}-\frac{1}{D_S}\biggr)\qquad \rm
\label{eqn:D_L}
$$

The resulting characteristics of the lens are given in Table \ref{table:tab3} for each model that we have presented : parallax only, parallax + xallarap (black-hole companion) and parallax + xallarap (solar-mass companion).
Due to the high parallax magnitude obtained with the ``black hole'' model (see Table \ref{table:tab1}), the lens mass is a brown dwarf ($M=0.025 M_{\rm{\odot}}$) in the extreme foreground ($D_L=824 pc$). Moreover, the extreme black hole mass ($M_s > 70M_{\rm{\odot}}$), by itself, virtually rules out this model. We take this as evidence for unrecognized systematic errors at the $\Delta \chi^2\sim 40$ level, and hence do not believe inferences based on $\Delta \chi^2$ at this level are robust. Systematic errors at this level are not uncommon for microlensing events.

The model with a solar-mass companion is suspect as well, still with a brown-dwarf lens in the foreground, meaning that it results from the same systematics.
We therefore conclude that the xallarap ``signal'' is probably spurious and we present these two models only for completeness. We expect that the presence of these systematics will corrupt the parallax measurements by of order $\sqrt{\Delta\chi^2_{xallarap+parallax}/\Delta\chi^2_{parallax}}\sim \sqrt{(235\pm43/235)}-1 \sim 9\%$, which will impact the lens mass and relative parallax estimates. However, this systematic error is too small to qualitatively impact the conclusions of this paper.

For the model with parallax effects only, the lens star is a M-dwarf (Table \ref{table:tab3}) and situated in the disk, lying $5.5$ \rm{kpc} from the observer. With the added uncertainties due to systematics, the parallax becomes $\pi_{\rm_E}=0.12\pm0.03\pm0.01$, the lens mass estimates $M=0.50\pm0.14\,M_{\rm{\odot}}$ (± 28\%) and the relative parallax $\pi_{rel}=57.9\pm14.5\mu \rm{as}$.
For the rest of the analysis, we will only consider this model when the physical parameters of the lens are needed.
\small{
\begin{table}
\begin{center}
\begin{tabular}{l r r r}
\multicolumn{4}{c}{TABLE 3} \\
\multicolumn{4}{c}{OGLE-2007-BLG-050   \textsc{Lens mass and distance}} \\
\multicolumn{4}{c}{Extended-Source Point-Lens with Parallax} \\
\hline
\hline
\multicolumn{1}{c}{Parameters} & \multicolumn{1}{c}{Without} & \multicolumn{1}{c}{Xallarap} &  \multicolumn{1}{c}{Xallarap}\\
\multicolumn{1}{c}{   } & \multicolumn{1}{c}{xallarap} & \multicolumn{1}{c}{Black hole} & \multicolumn{1}{c}{Solar mass}\\
\multicolumn{1}{c}{   } & \multicolumn{1}{c}{   } & \multicolumn{1}{c}{P = 290 days} & \multicolumn{1}{c}{P = 170 days}\\
\hline
\\
$\theta_{\rm E}$  & $0.48\pm0.01$ & $0.47\pm0.041$ & $0.47\pm0.01$\\
$M$ \textit{   ($M_{\rm{\odot}}$)}& $0.50\pm0.13$ & $0.025\pm0.001$ & $0.0618\pm0.0007$\\
$\pi_{\rm rel}$ \textit{   ($\mu \rm{as}$)} & $58\pm15$ & $1088\pm46$ & $440\pm58$\\
$D_L$    \textit{   (kpc)} & $5.47\pm0.45$ & $0.82\pm0.07$ & $1.77\pm0.20$ \\
\hline
\end{tabular}
\label{tab3}
\end{center}
\caption{\label{table:tab3}
Lens mass and distance for three different models : 1/ with orbital parallax effects only, 2/ with orbital parallax + xallarap (black-hole source companion), 3/ with orbital parallax + xallarap (solar-mass source companion).
}
\end{table}
}

\normalsize
As discussed by \cite{ghosh04}, future high-resolution astrometry could allow the direct measurement of the magnitude and direction of the lens-source relative proper motion $\bmu$ and substantially reduce the parallax uncertainty and thus the stellar mass uncertainty. But according to our initial estimate of the relative proper motion ($\mu=2.63\pm0.08\,$mas/yr), it would take at least a 20 years to clearly detect the lens (especially since the source is very bright), but hopefully, within a decade, either ELT, GMT or TMT (giant telescopes) will be built, in which case the lens could be observed thereafter.

\section{Planet Detection efficiency}

\subsection{Introduction and previous analyses}

To provide reliable abundance limits of Jupiter- to Earth-mass planets in our Galaxy, it is essential to evaluate the apparent non-planetary events, especially the well-covered high magnification events. A necessary step is to evaluate the confidence with which one can exclude potential planetary companions for each event.

Since OGLE-2007-BLG-050 presents strong finite-source effects, one may wonder whether a given planetary perturbation would have been so washed out by these effects as to become undetectable. Using many such efficiency calculations the aim is to determine the selection function to the underlying population of planets.

\cite{gaudisack00} developed the first method to calculate detection efficiency for a single planet, which was extended to multiple planets detection efficiency by \cite{gaudi02}, who analyzed 43 microlensing events from the 1995-1999 observational seasons. Three of them were high magnification events [OGLE-1998-BLG-15 ($A_{max}\sim 170$), MACHO-1998-BLG-35 ($A_{max}\sim 100$) and OGLE-1999-BLG-35 ($A_{max}\sim 125$)]. This 5-year analysis provided the first significant upper abundance limit of Jupiter- and Saturn-mass planets around M-dwarfs. \cite{tsapras03} and \cite{snodgrass04} derived constraints on Jovian planet abundance based on OGLE survey data of 1998-2000 and 2002 seasons respectively.

Computing detection efficiency for individual events is thus required to estimate the frequency of planetary signatures in microlensing light curves, and a couple of complex events have indeed been analyzed separately. For example the high magnification event OGLE-2003-BLG-423 ($A_{max}\sim  256$) by \cite{yoo04b} who found that the event was not as sensitive as it should have been if better monitored over the peak. Another high magnification ($A_{max}\sim 525$) example is MOA-2003-BLG-32 / OGLE-2003-BLG-219 was analyzed by \cite{abe04} and \cite{dong06} (Appendix B). This well-covered event showed the best sensitivity to low-mass planets to date. Finally, the highest magnification event ever analyzed, OGLE-2004-BLG-343, was unfortunately poorly monitored over its peak, and \cite{dong06} showed that it otherwise would have been extremely sensitive to low-mass planets.

\subsection{Planet detection efficiency in Einstein Units}

To characterize the planetary detection efficiency of OGLE-2007-BLG-050, we follow the \cite{gaudisack00} method which consists of fitting binary models with the 3 binary parameters $(d,q,\alpha)$ held fixed and the single lens parameters allowed to vary. Here $d$ is the planet-star separation in units of $\theta_{\rm E}$, $q$ the planet-lens mass ratio, and $\alpha$ the angle of the source trajectory relative to the binary axis. In \cite{gaudisack00}, the single lens parameters, $u_0$, $t_0$ and $t_{\rm E}$, are related to a PSPL fit. In this analysis, we also fit the radius of the source $\rho_*$ (scaled to the Einstein radius) and compare the binary lens fits to the best ESPL fit for this event.

From the resulting fitted binary lens $\chi^2_{(d,q,\alpha)}$, we calculate the $\chi^2$ improvement : $\Delta\chi^2_{(d,q,\alpha)}=\chi^2_{(d,q,\alpha)}-\chi^2_{ESPL}$, and $\Delta\chi^2_{(d,q,\alpha)}$ is compared with a threshold value $\chi^2_C$. If $\Delta\chi^2_{(d,q,\alpha)}<-\chi^2_C$, the $(d,q,\alpha)$ planetary (or binary) system is detected, while if $\Delta\chi^2_{(d,q,\alpha)}>\chi^2_C$, it is excluded. \cite{gaudi02} argued that a threshold of 60 is high enough to be confident in excluding binary lens systems.

For each $(d,q)$, the fraction of angles $0<\alpha<2\pi$ that was excluded is called the ''sensitivity'' for that system. Indeed, the detection efficiency $\epsilon(d,q)$ can be expressed as :
$$\epsilon(d,q)=\frac{1}{2\pi}\int_0^{2\pi}{\Theta(\Delta\chi^2(d,q,\alpha)-\chi^2_C)d\alpha} 
\label{eqn:efficiency}
$$
where $\theta$ is the step function.
To perform the calculations of binary light curves, we use a binary-lens finite-source algorithm developed by \cite{dong06} (Appendix A). The resulting grids of $\chi^2$ as a function of $d$ and $\alpha$ are shown in Fig.\ref{fig:diag1} for some values of $q$. The complete computation has been done for every possible combinations between the following values of $d$, $q$ and $\alpha$ :

\begin{itemize}
 \item $q$ : 19 values with a constant logarithmic step over the range [$10^{-6}$, $10^{-2}$].
\item $d$ : 40 values with a constant logarithmic step over the range [$0.1$, $10$].
\item $\alpha$ : 121 values linearly spaced from $0$ to $360^o$.

\end{itemize}

The resulting detection efficiency diagram for OGLE-2007-BLG-050 is shown in Fig.\ref{fig:diag2}. The first observation is that no planet is detected since there is no configuration that gives $\Delta\chi^2 < -60$. This event is very sensitive to the presence of planets, especially in the [0.8 - 1.2] separation range in Einstein units, where the detection efficiency reaches 100\% for Jupiter mass ratios ($q=9x10^{-4}$), 75\% for Neptune mass ratios ($q=5x10^{-5}$) and 10\% for Earth mass ratios ($3x10^{-6}$). In larger separation ranges, as [0.4 - 2.7] $R_{\rm E}$, we exclude Jupiter mass ratios with 95\% confidence.

In future statistical analyses of microlensing planetary detection efficiency, one will likely be forced to use a higher exclusion threshold than 60 because, while planets can sometimes be reliably excluded at this threshold (as in the present case), it is unlikely that they can be reliably detected at this level, particularly in high-magnification events. Because we cannot predict the exact threshold that will be adopted by future studies, we show both our exclusion level ($\Delta\chi^2 > 60$) and a somewhat arbitrarily chosen value, $\Delta\chi^2 > 250$. The important point is that the detection efficiency diagrams in the two cases (Fig.\ref{fig:diag2} and with a threshold equal to 250 in Fig.\ref{fig:diag250}) are very similar.

\subsection{Planet detection efficiency in physical units}

Having an estimate of the angular Einstein radius $\theta_{\rm E}$, the distance $D_L$
of the lens from the observer and the lens mass $M$, we derive estimates of the
physical parameters $(r_\perp,m_p)$ for the tested planetary models, where $r_\perp$
is the projected separation between the planet and its host star and $m_p$ the planet
mass, and calculate the associated detection efficiency.
$$r_\perp\,({\rm AU})=d\,D_L\,({\rm kpc})\,\theta_{\rm E}\,({\rm mas})
\label{eqn:radius}
$$
$$m_p=qM
\label{eqn:planetmass}
$$

To simplify the translation between efficiency diagrams in Einstein units and physical units, we have considered the values of $M$, $D_L$ and $\theta_{\rm E}$ as perfectly known. A proper analysis would involve a convolution of the detection efficiency map in terms of native parameters $d,q$ over the probability density distribution of the primary lens parameters (e.g. \cite{yoo04b}). While our procedure of keeping $M$, $D_L$ and $\theta_{\rm_E}$ fixed is an approximation, considering the logarithmic scale of the efficiency maps, the uncertainties in the primary lens parameters will not have an important effect on the shape of the resulting efficiency diagrams. 

We take the parameters related to the fit with extended source and parallax effects, where $M=0.50 \pm 0.14\, M_{\rm{\odot}}$, $D_L=5.47\pm0.45\,kpc$ and $\theta_{\rm_E}=0.48\pm0.01\,mas$.
The resulting detection efficiency diagram in physical units is shown in Fig.\ref{fig:diag2} as well, but the corresponding axes are those on the top and the right of the graphic. This demonstrates that OGLE-2007-BLG-050 is sensitive to Neptune-mass planets as well as some Earth-mass configurations. Indeed, for a [1.8 - 3.1] AU projected separation range between the planet and the lens star, Jupiter, Neptune and Earth-like planets are excluded with a 100\%, 95\% and 10\% confidence respectively. For a range of [1.4 - 4] AU, the detection efficiency reaches 100\% for Jupiter mass planets and 75\% for Neptune mass planets, and for a much bigger range of [0.6 - 10] AU, Jupiter-like planets are excluded with a 75\% confidence.

\subsection{Planet detection efficiency as a function of central caustic size}

\cite{chung05} analyzed the properties of central caustics in planetary microlensing events in order to estimate the perturbation that they induce. They gave an expression for the central-caustic size as a function of the planet-star separation and the planet/star mass ratio. Several authors have considered the size and shape of the central caustic as a function of the parameters of the planet for high-magnification events \citep{griestsaf98,domi99,dong08b}.
In the analysis of the cool Jovian-mass planet MOA-2007-BLG-400Lb, \cite{dong08b} conducted the initial parameter space search over a grid of ($w$,$q$) rather than ($d$,$q$) where $w$ is the ``width'' of the central caustic. For MOA-2007-BLG-400, the angular size of the central caustic is smaller than that of the source ($w/\rho \sim0.4$), and $w$ can be directly estimated by inspecting the light curve features. \cite{dong08b} find the ($w$,$q$) parametrization is more regularly defined and more efficient in searching parameter space than ($d$,$q$).

The source size of OGLE-2007-BLG-050 is $\rho = 0.0045$ which is relatively big, and since finite-source effects smear out the sharp magnification pattern produced by the central caustics, one way to present the planetary detection efficiency results is to estimate the ratio $w/\rho$ that is reached at the detection/exclusion limits. 
Assuming that detectable planets should produce signals $\geq 5\%$, \cite{han09} estimated the ratio $w/\rho$ must be at least equal to 0.25.
Here we present the detection efficiency diagram in ($d$,$w/\rho$) space in Fig.\ref{fig:diag3}, still considering $\Delta\chi^2>60$ as the criterion of exclusion. This diagram shows a clear frontier in red at $w/\rho$ values between 0.1 and 0.3 above which the detection efficiency is greater than $50\%$, which also corresponds to the 50\% detection's contours in Fig.\ref{fig:diag2}. On this frontier, the value of $w/\rho$ goes down to 0.1 for $d\sim1$ and increases to 0.3 for $d>>1$ or $d<<1$. Our realistic estimate of detection efficiency is in general agreement with the simple criterion in \cite{han09}. Given the high photometric precision and dense sampling, our data allow detections below the 5\% threshold adopted by \cite{han09}. We also note that the $w/\rho$ threshold is weakly dependant on $d$, which is a result of the enhancement in detection efficiency of the resonant caustics at small mass ratios. 

We have presented a new way of visualizing the detection efficiency in ($d$,$w/\rho$) space. It offers a physically straightforward way to understand the planetary sensitivity in events with pronounced finite-source effects. We find that the data obtained by current observation campaigns can probe planetary central caustics as small as $\sim 20\%$ of the source size for high-magnification events.

\section{Conclusion}

OGLE-2007-BLG-050 is a rare case of a high magnification event with well measured finite source effects and detectable parallax effects. This leads to an estimate of the angular Einstein radius $\theta_{\rm E}=0.48\pm0.01$ mas, the parallax $\pi_{\rm E}=0.12\pm0.03$, the mass $M=0.50\pm0.14\,M_{\odot}$ and distance $D_L=5.5\pm0.4$ kpc of the lens star. This is only the second reasonably precise mass estimate (to within 28\%) for an unseen single object using any method.

When computing planet detection efficiency, we did not find any planetary signature and the resulting maps in $(d,q,\alpha)$, where $d$ is the planet-star separation in Einstein units, $q$ the planet-lens mass ratio, and $\alpha$ the angle of the source trajectory relative to the binary axis, reveal a good sensitivity to low mass ratios $q$, with a 75\% and 10\% efficiencies for Neptune- and Earth-mass ratios respectively in the range [0.8 - 1.2] $R_{\rm{E}}$, and a 100\% detection efficiency for Jupiter-mass ratio in [0.4 - 2.7] $R_{\rm{E}}$.

It also permits the calculation of efficiency maps in physical space $(r_\perp,m_p)$, where $r_\perp$ is the projected planet/star separation and $m_p$ is the planet mass. Here we show that this microlensing event is very sensitive to Neptune-mass planets and has (10\%) sensitivity to Earth-mass planets within a [1.8 - 3.1] AU projected separation range.

\begin{acknowledgements}
We thank Thomas Prado and Arnaud Tribolet for their careful reading of the manuscript.
VB thanks Ohio State University for its hospitality during a six week visit, during which this study was initiated. 
We acknowledge the following support:
Grants HOLMES ANR-06-BLAN-0416
Dave Warren for the Mt Canopus Observatory;
NSF AST-0757888 (AG,SD); NASA NNG04GL51G (DD,AG,RP);
Polish MNiSW N20303032/4275 (AU);
HST-GO-11311 (KS);
NSF AST-0206189 and AST-0708890, NASA NAF5-13042 and NNX07AL71G (DPB);
Korea Science and Engineering Foundation grant 2009-008561 (CH);
Korea Research Foundation grant 2006-311-C00072 (B-GP);
Korea Astronomy and Space Science Institute (KASI);
Deutsche Forschungsgemeinschaft (CSB);
PPARC/STFC, EU FP6 programme ``ANGLES'' ({\L}W,NJR);
PPARC/STFC (RoboNet);
Dill Faulkes Educational Trust (Faulkes Telescope North);
Grants JSPS18253002, JSPS20340052 and JSPS19340058 (MOA);
Marsden Fund of NZ(IAB, PCMY); 
Foundation for Research Science and Technology of NZ;
Creative Research Initiative program (2009-008561) (CH);
Grants MEXT19015005 and JSPS18749004 (TS).
This work was supported in part by an allocation of computing time from the Ohio Supercomputer Center.
\end{acknowledgements}

\onecolumn{
\begin{figure}
\includegraphics[width=2.3in]{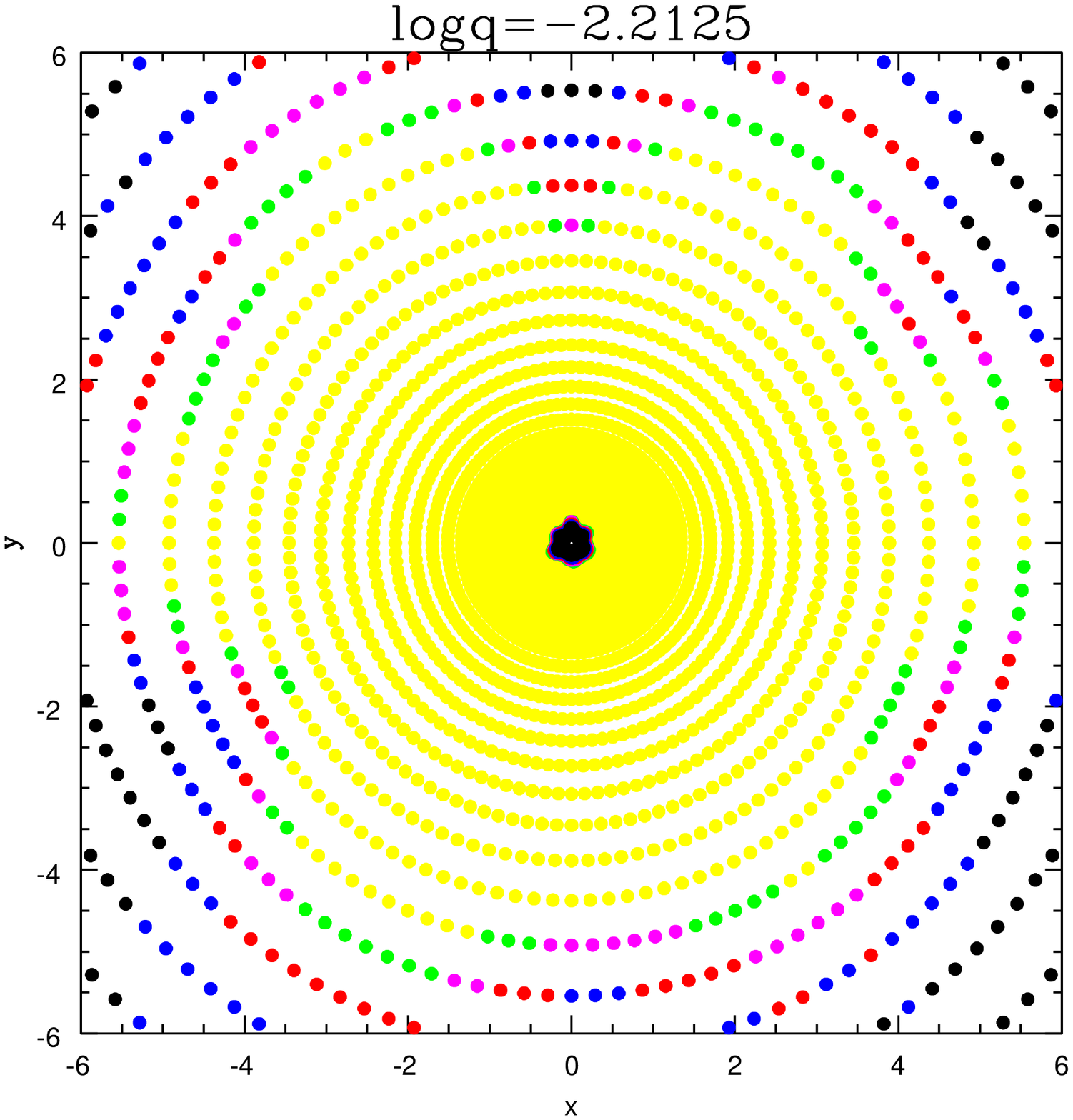}
\includegraphics[width=2.3in]{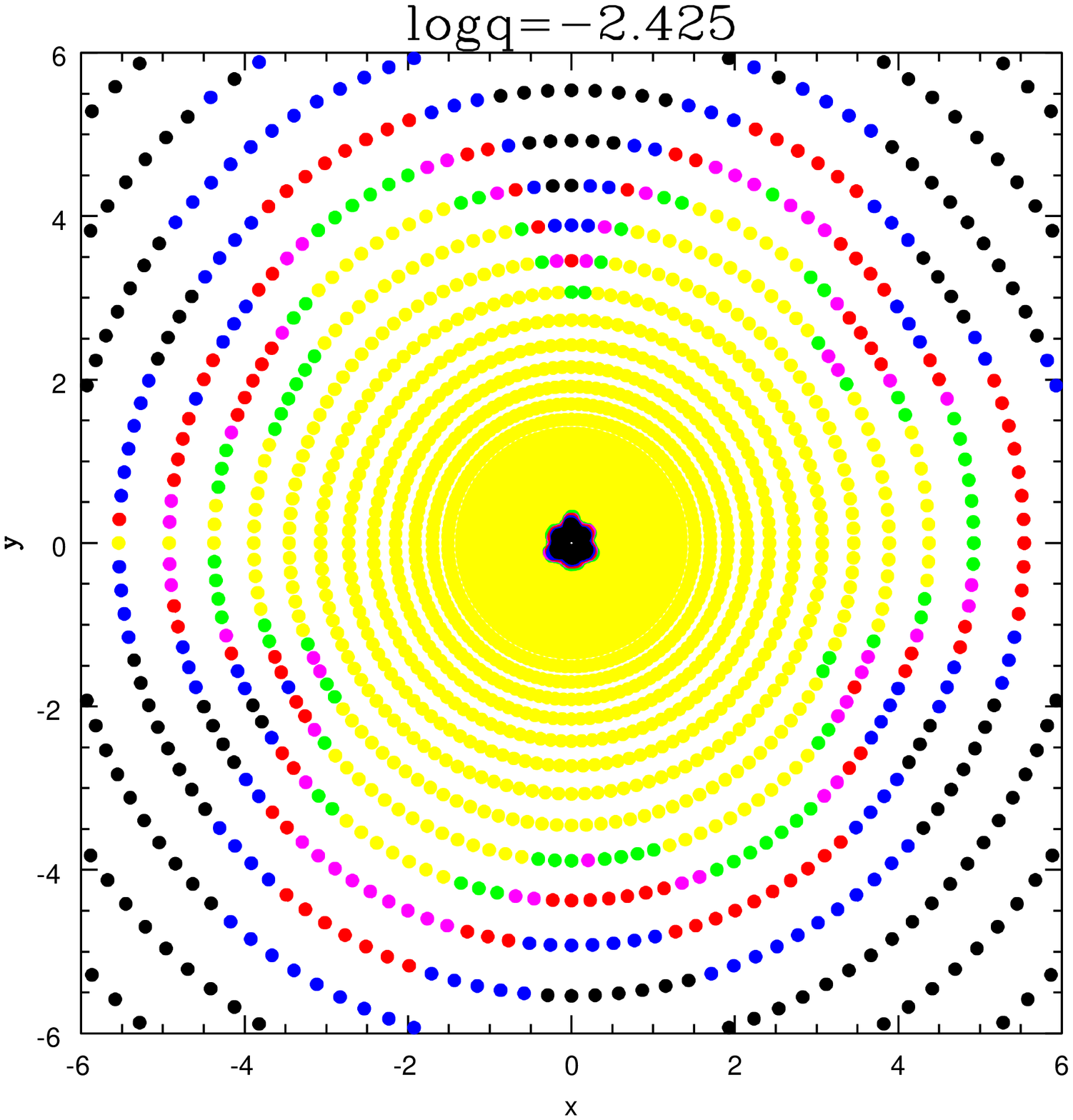}
\includegraphics[width=2.3in]{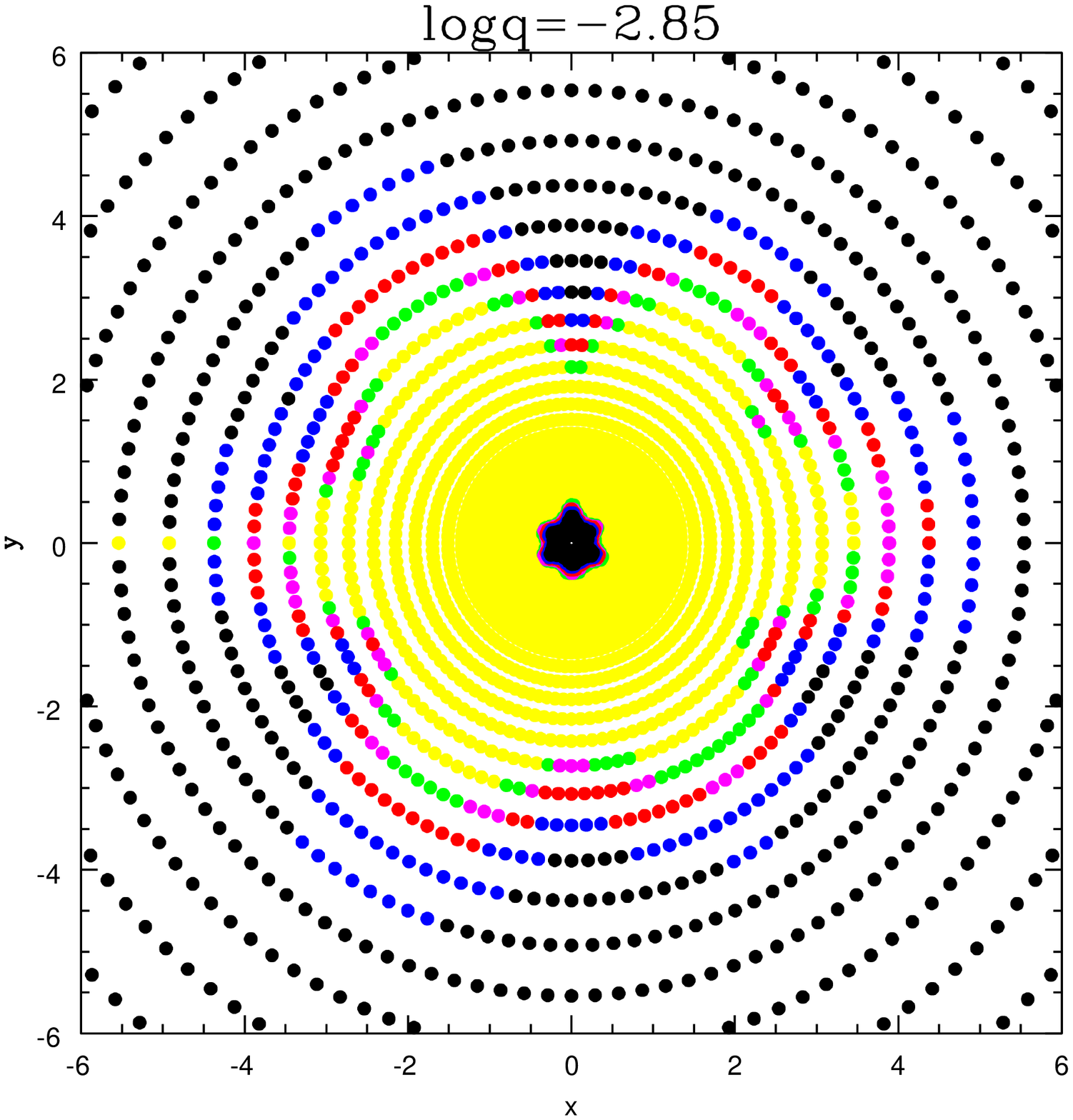}
\includegraphics[width=2.3in]{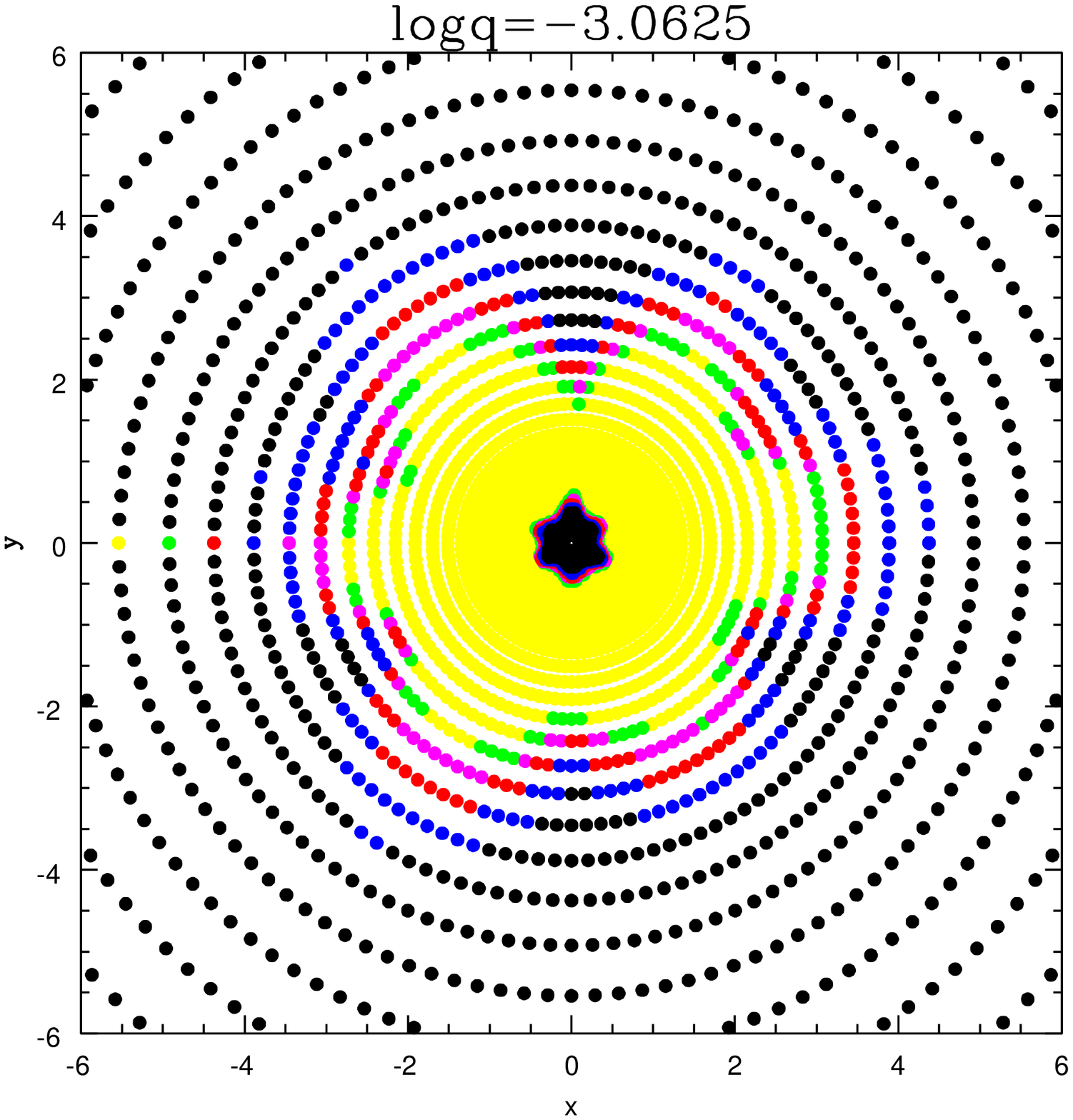}
\includegraphics[width=2.3in]{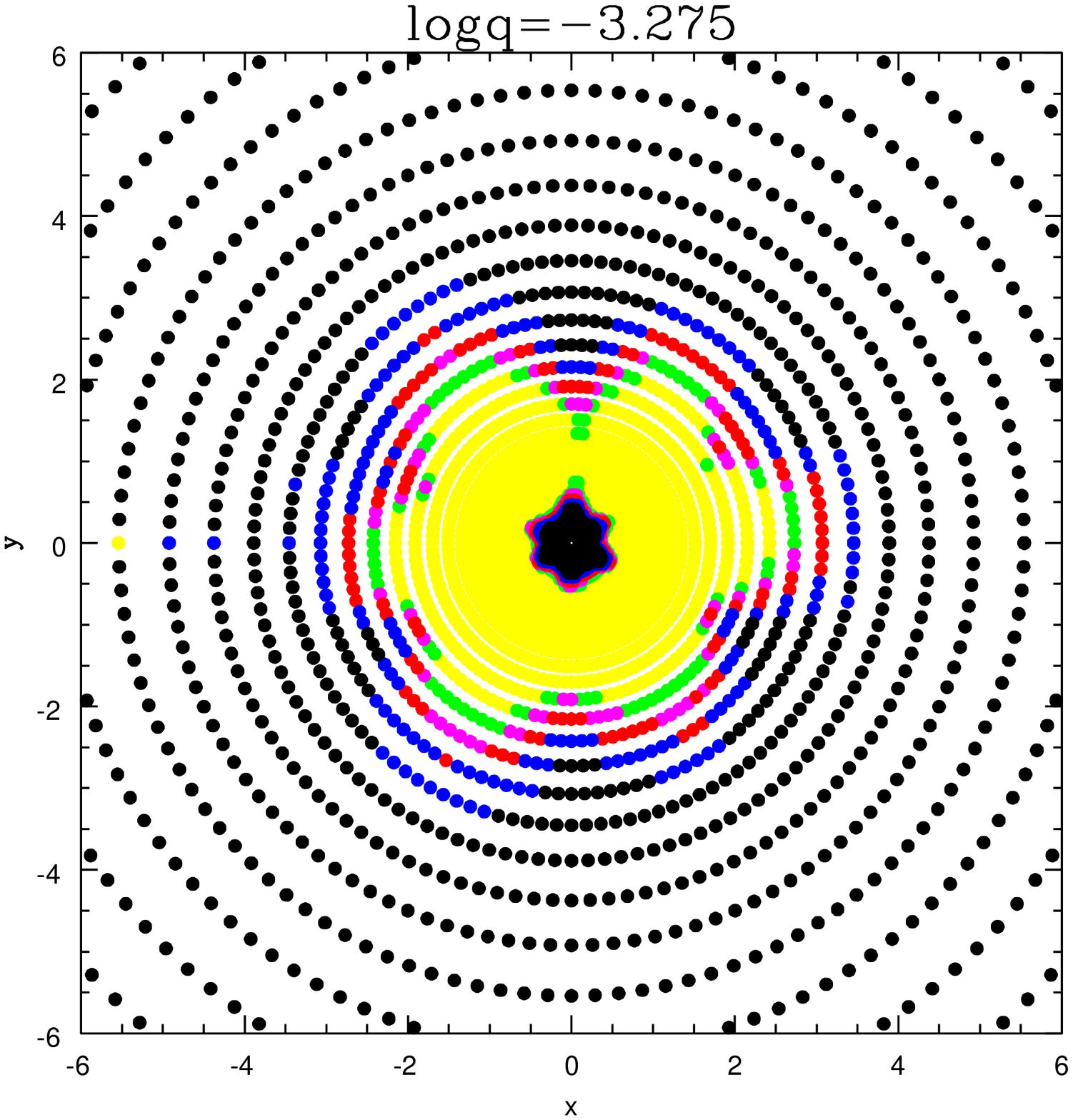}
\includegraphics[width=2.3in]{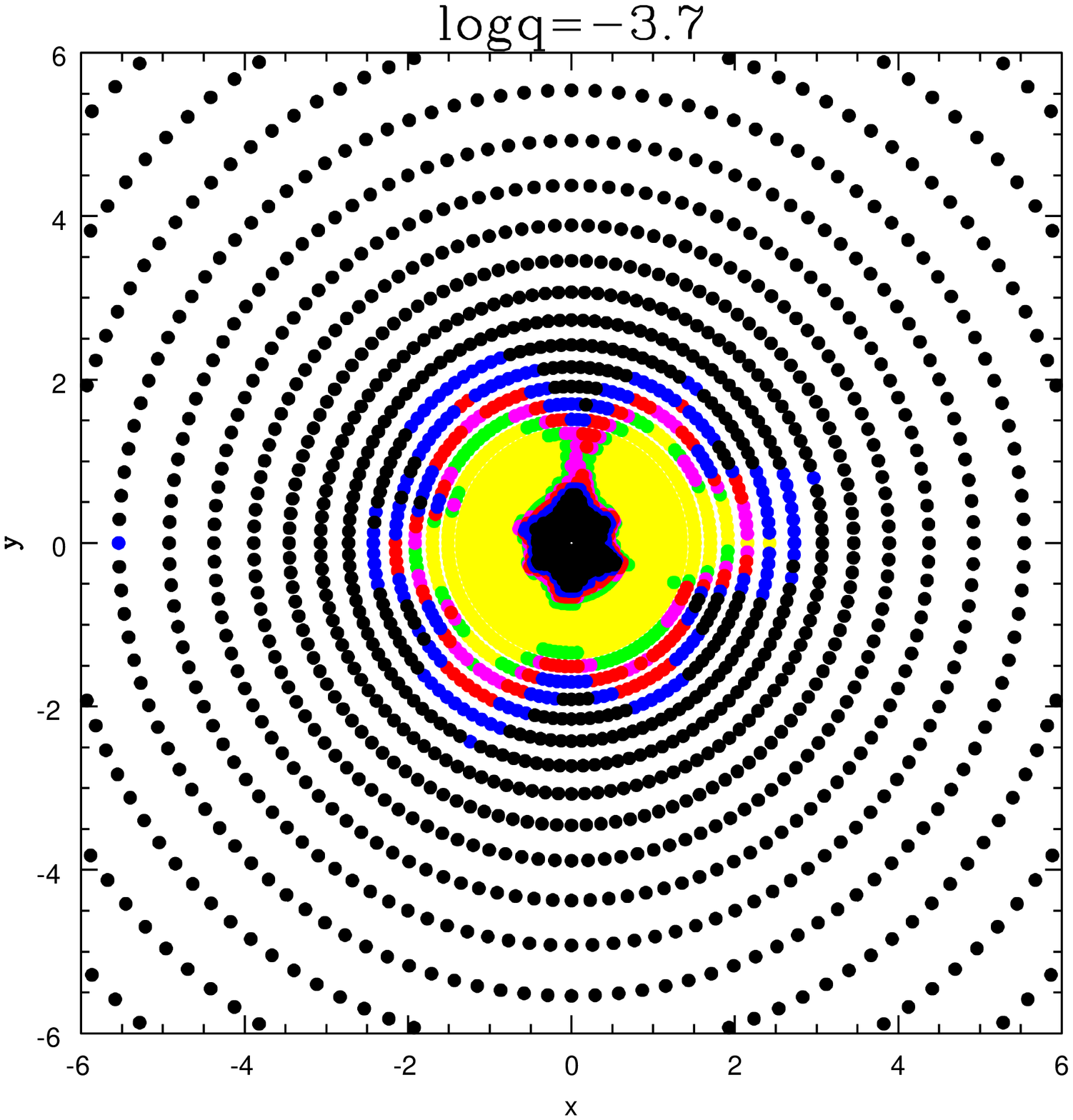}
\includegraphics[width=2.3in]{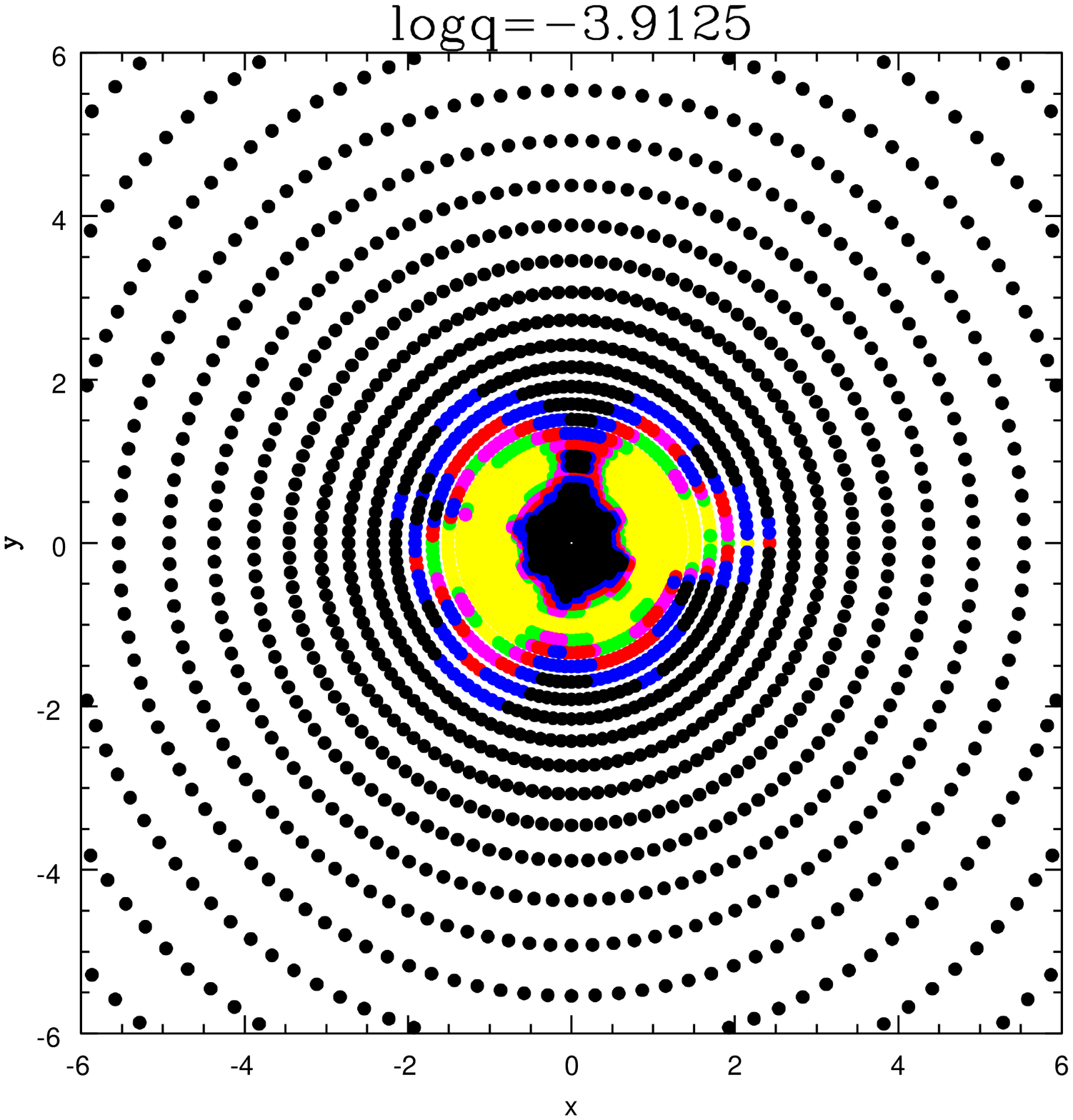}
\includegraphics[width=2.3in]{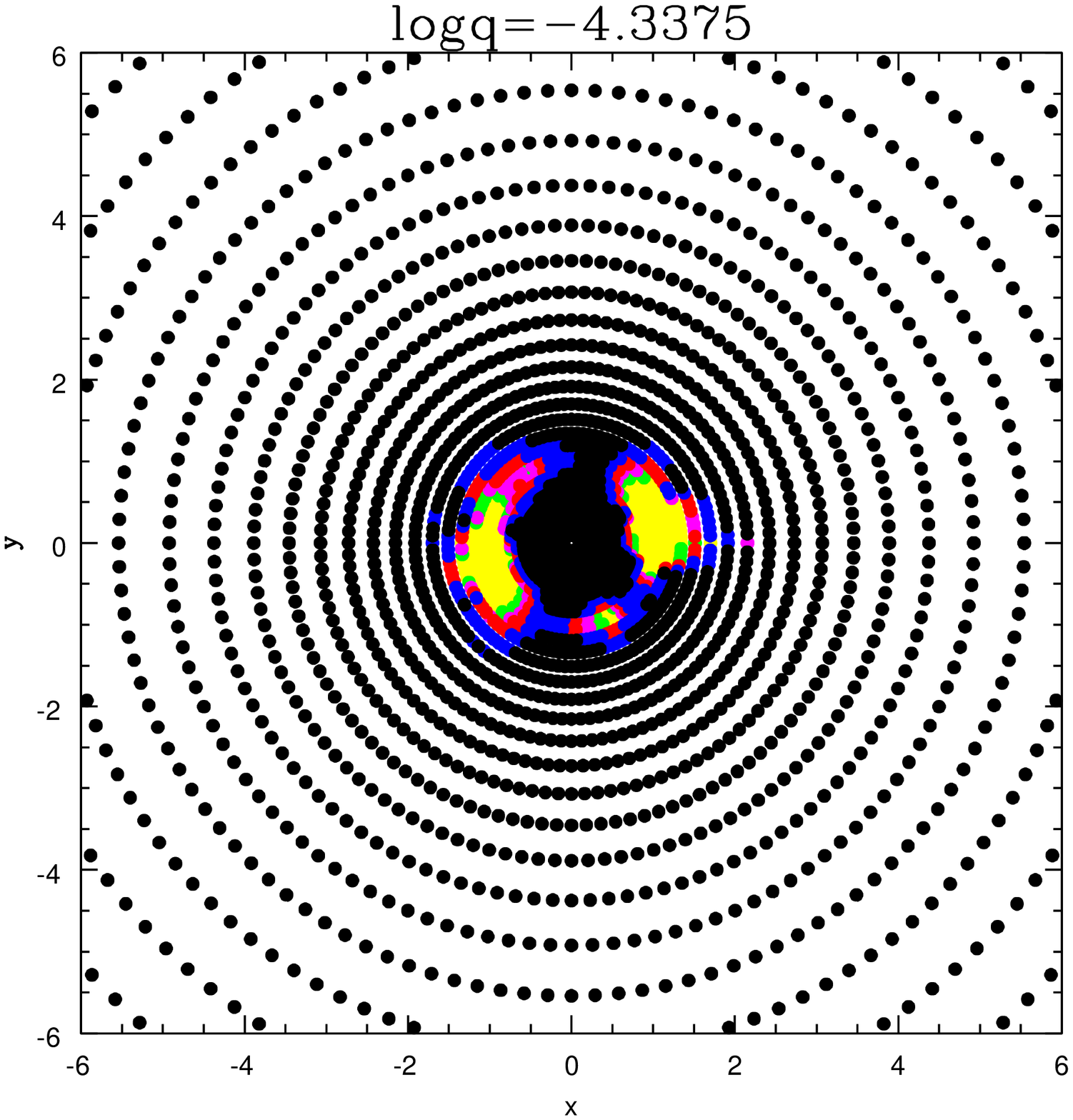}
\includegraphics[width=2.3in]{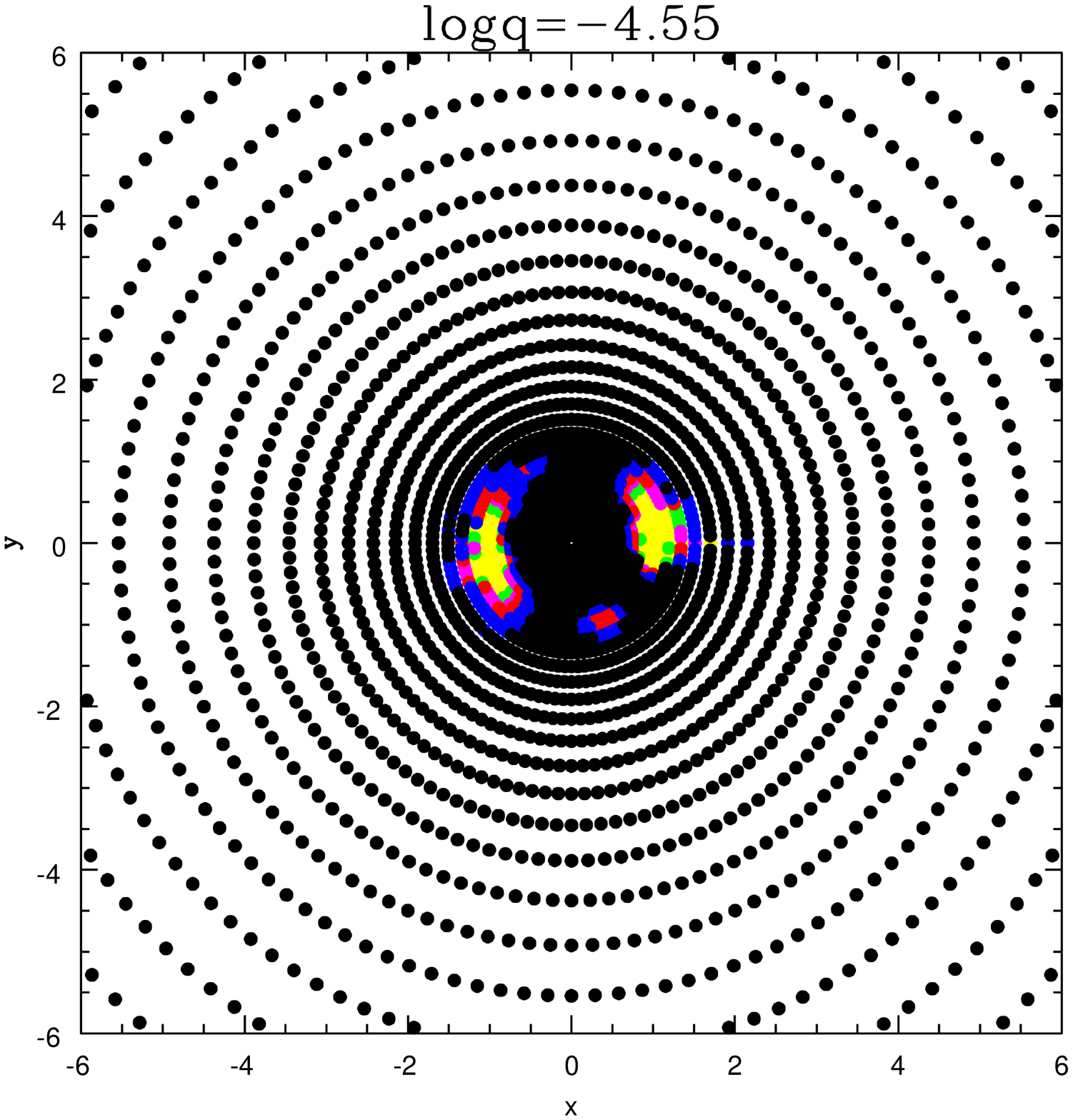}
\includegraphics[width=2.3in]{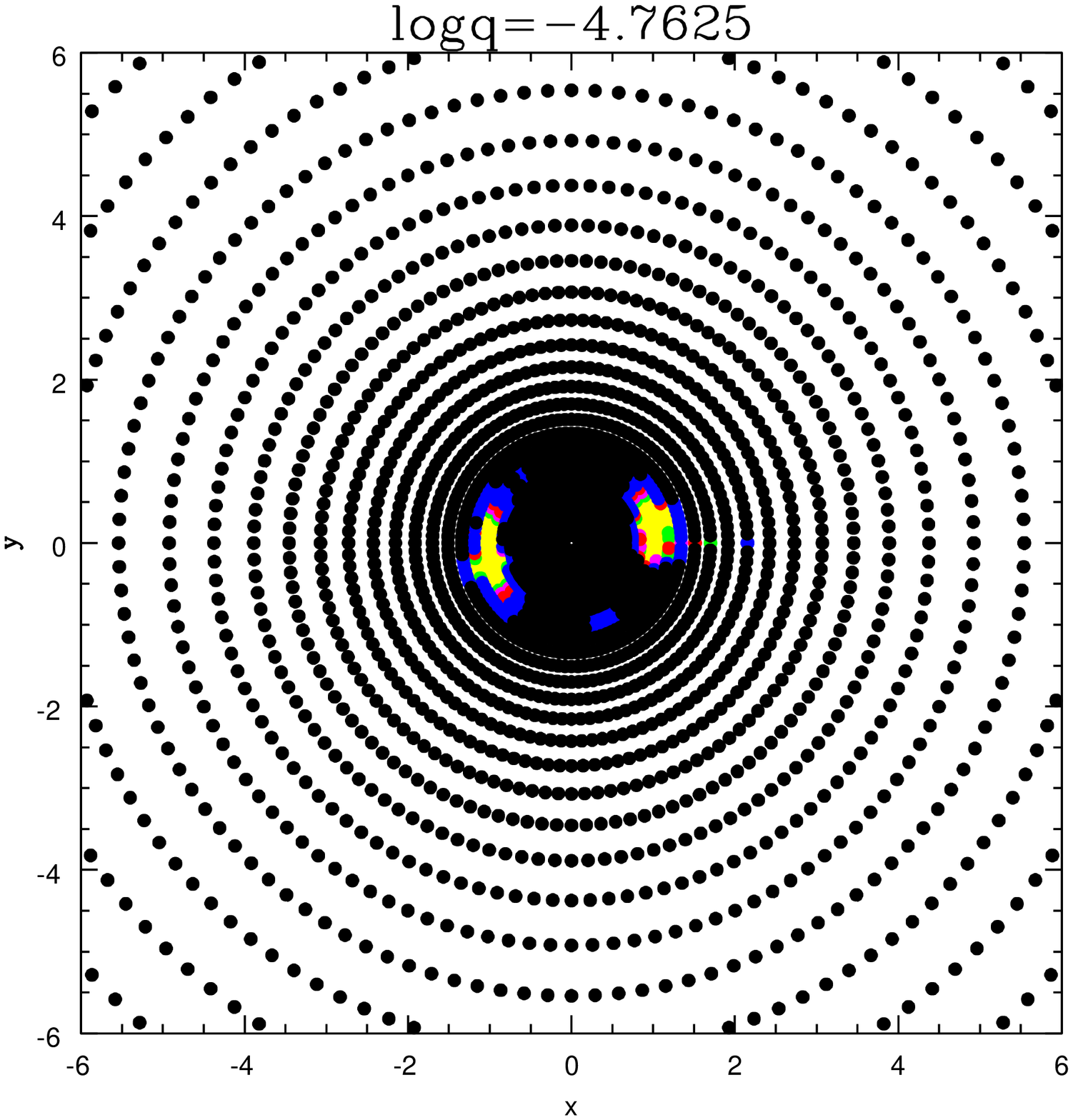}
\hspace{0.1cm}
\includegraphics[width=2.3in]{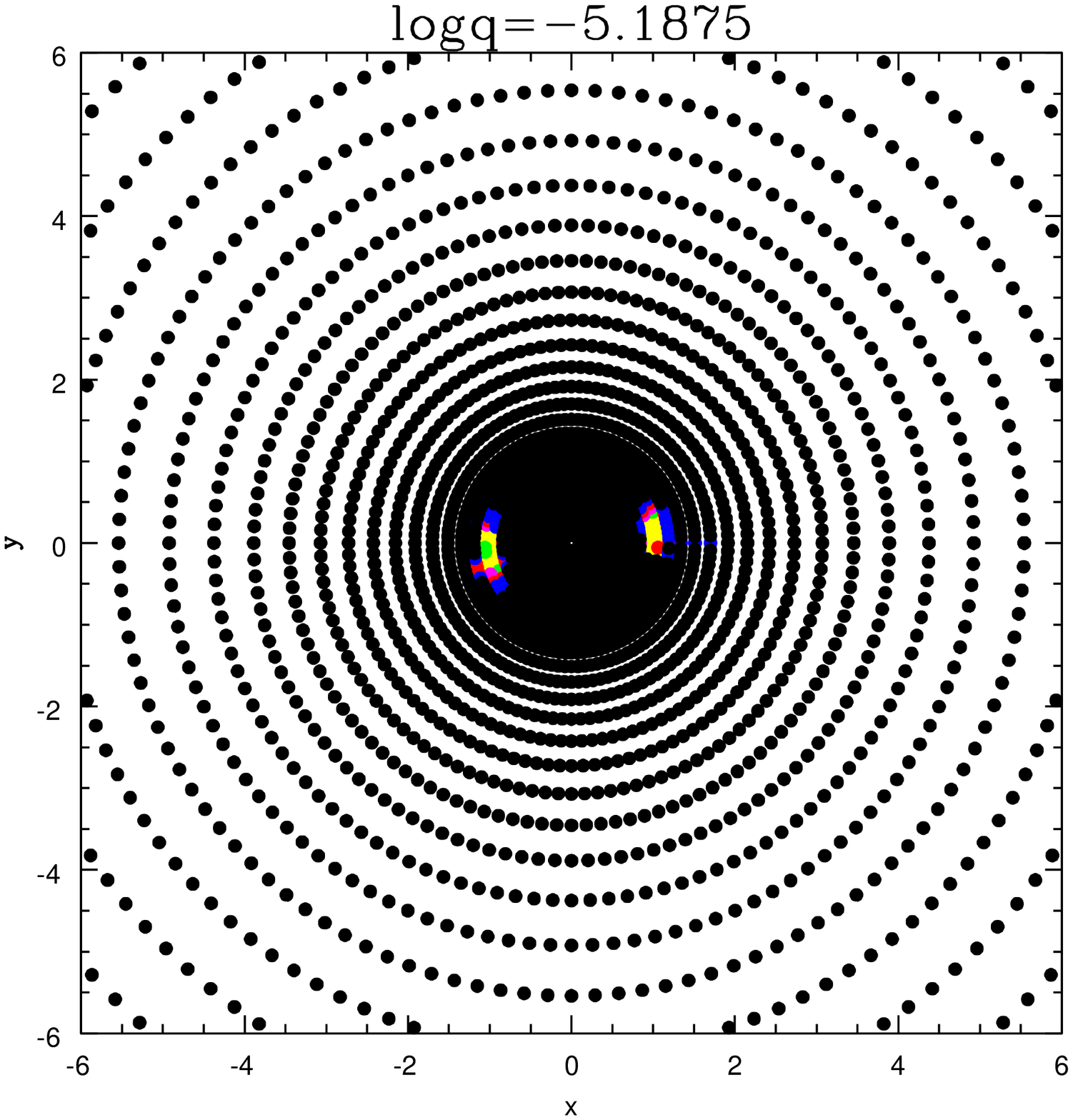}
\hspace{0.1cm}
\includegraphics[width=2.3in]{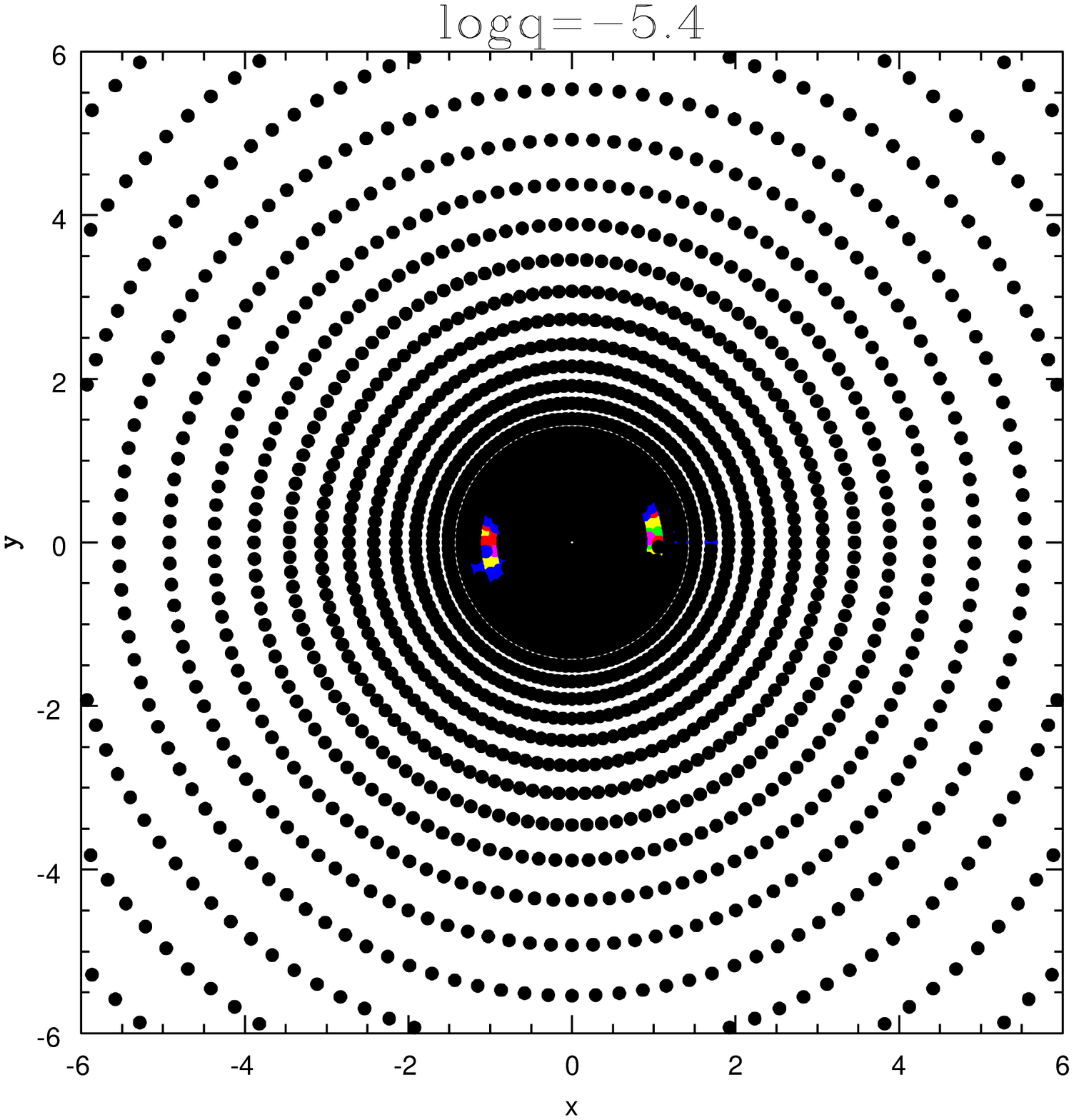}
\caption{\label{fig:diag1}
Binary-lens finite-source grids of $\chi^2$ as a function of $(x,y)$ where $x=d\cos{\alpha}$ and $y=d\sin{\alpha}$ for different values of $q$. The value appearing in the upper part of each diagram corresponds to the value of $\log{q}$. The color scale shows the variations of the resulting $\Delta\chi^2$, where $\Delta\chi^2$ is the difference between a given binary lens model $\chi^2$ and the ESPL fit $\chi^2$, $\Delta\chi^2_{(d,q,\alpha)}=\chi^2_{(d,q,\alpha)}-\chi^2_{ESPL}$. The colors correspond to the following thresholds : {black, blue, red, magenta, green, yellow} = {$\Delta\chi^2<60$, 60-100, 100-150, 150-200, 200-300, $300<\Delta\chi^2$}. These diagrams have been computed for $d$ and $q$ ranges of $[0.1-10]R_{\rm E}$ and $[10^{-6}-10^{-2}]M_{lens}$ and for 121 values of $\alpha$.
}\end{figure}

}

\onecolumn{
\begin{figure}
\begin{center}
\includegraphics[width=4in]{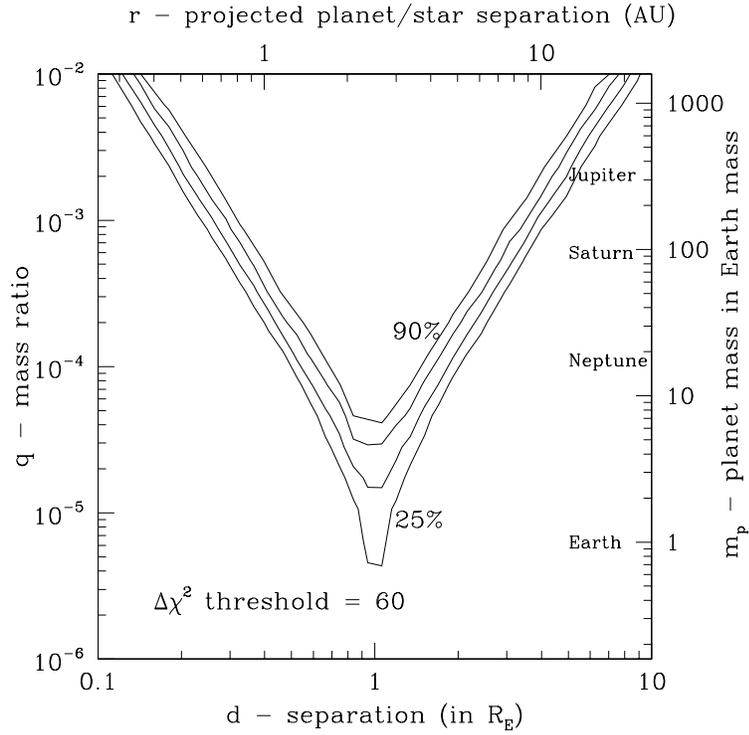}
\caption{\label{fig:diag2}
Resulting detection efficiency diagram for $d$ and $q$ ranges of $[0.1-10]R_{\rm E}$ and $[10^{-6}-10^{-2}]M_{lens}$ and detection efficiency diagram in physical units $(r_\perp,m_p)$ if considering the upper and right axes. The contours indicate 25\%, 50\%, 75\% and 90\% efficiency, with an excluding threshold equal to 60.
}
\end{center}
\end{figure}

\begin{figure}
\begin{center}
\includegraphics[width=4in]{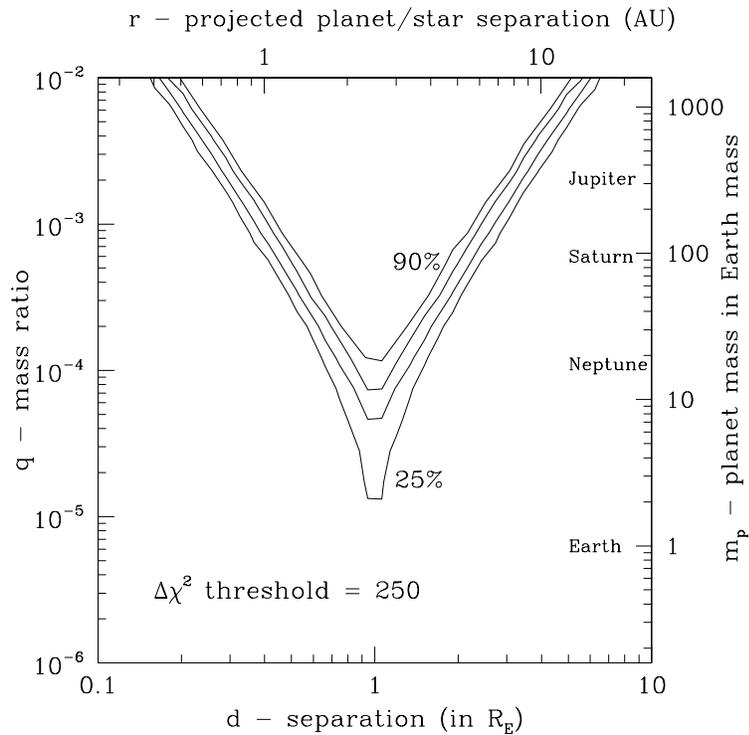}
\caption{\label{fig:diag250}
Same as Fig.\ref{fig:diag2}, except with a hypothetical threshold of $\Delta\chi^2=250$. Comparison with Fig.\ref{fig:diag2} shows that planet sensitivity does not depend strongly on threshold.
}
\end{center}
\end{figure}

\begin{figure}
\begin{center}
\includegraphics[width=3in]{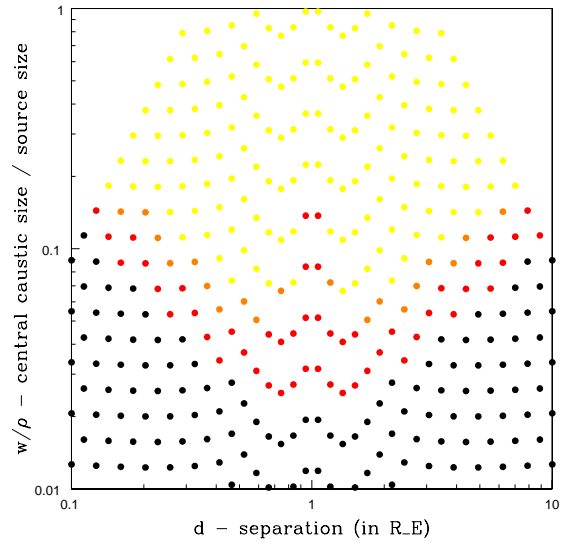}
\caption{\label{fig:diag3}
Resulting detection efficiency diagram in ($d$,$w/\rho$) space. This diagram shows a clear frontier in red at $w/\rho$ values between 0.1 and 0.3 above which the detection efficiency is greater than $50\%$. This frontier corresponds to the 50\% detection's contours in the Fig.\ref{fig:diag2}.
}
\end{center}
\end{figure}

}

\onecolumn{



}
}
\end{document}